\documentclass[showpacs,showkeys,twocolumn,pra,superscriptaddress]{revtex4-1}%
\pdfoutput=1
\usepackage{graphics,amsmath,amsfonts,amscd,revsymb,latexsym,
enumerate,multirow,epsfig}
\usepackage{times}
\usepackage{mathtools}
\usepackage[caption=false]{subfig}
\usepackage{amssymb}
\usepackage{xcolor}
\usepackage{graphicx}%
\usepackage{bbm}
\usepackage{slashed}
\usepackage{tabularx}
\usepackage{tablefootnote}
\usepackage[capitalise]{cleveref}

\providecommand{\U}[1]{\protect\rule{.1in}{.1in}}
\providecommand{\U}[1]{\protect\rule{.1in}{.1in}}

\begin{document}

\title{Fermion-bag inspired Hamiltonian lattice field theory for fermionic quantum criticality}

\author{Emilie Huffman}
\affiliation{Perimeter Institute for Theoretical Physics}
\author{Shailesh Chandrasekharan}
\affiliation{Duke University}
\begin{abstract}

Motivated by the fermion bag approach we construct a new class of Hamiltonian lattice field theories that can help us to study fermionic quantum critical points, particularly those with four-fermion interactions. Although these theories are constructed in discrete-time with a finite temporal lattice spacing $\varepsilon$, when $\varepsilon\rightarrow 0$, conventional continuous-time Hamiltonian lattice field theories are recovered. The fermion bag algorithms run relatively faster when $\varepsilon=1$ as compared to $\varepsilon \rightarrow 0$, but still allow us to compute universal quantities near the quantum critical point even at such a large value of $\varepsilon$. As an example of this new approach, here we study the $N_f=1$ Gross-Neveu chiral Ising universality class in $2+1$ dimensions by calculating the critical scaling of the staggered mass order parameter. We show that we are able to study lattice sizes up to $100^2$ sites when $\varepsilon=1$, while with comparable resources we can only reach lattice sizes of up to $64^2$ when $\varepsilon \rightarrow 0$. The critical exponents obtained in both these studies match within errors.

\end{abstract}
\maketitle

\section{Introduction}
\label{sec:intro}

The effort to understand quantum critical points resulting from fermions that do not decouple at low energies and long distances is an exciting area of research across energy scales. In $2+1$ dimensions, it is well-known that relativistic four-fermion models containing massless Dirac fermions can exhibit the presence of such critical points \cite{Rosenstein:1990nm}. These four-fermion models are usually referred to as either Gross-Neveu models \cite{PhysRevD.10.3235} or Thirring models \cite{Thirring:1958in} depending on the type of interaction and have been studied extensively over the years \cite{ZinnJustin:1991yn,Hands:1992be,Ros93,Gracey:1992,Gracey:1993pd,Kark94,PhysRevD.53.4616,DelDebbio:1999he,PhysRevD.63.054502,Christofi:2006zt,PhysRevD.75.101701}. Study of quantum critical points in these four-fermion models has reemerged as an exciting area of research \cite{PhysRevD.92.085046}, especially due to the recent discovery that many materials can be described by Dirac fermions in the low energy limit and such materials can have new phases and quantum critical points that separate them \cite{Vafek:2013mpa,Topins2013}. Massless fermions can even help induce new quantum critical points and multi-critical points that do not exist in purely bosonic models \cite{PhysRevB.97.125137,PhysRevB.97.041117,PhysRevD.96.056004,Torres:2019vcw}. New analytical studies of the Gross-Neveu transitions using $\epsilon$-expansions \cite{Ihrig:2018hho,Zerf:2017zqi,PhysRevB.94.245102}, large-N expansions \cite{PhysRevD.97.105009,Gracey:2018ame}, functional renormalization group techniques \cite{PhysRevB.95.075101} and the bootstrap approach \cite{Iliesiu:2015qra} have been performed recently. This progress, combined with new solutions to fermion sign problems \cite{Chandrasekharan:2013rpa,PhysRevD.85.091502,PhysRevB.89.111101,Wang:2015vha,Li:2016gte,Wei:2016sgb,PhysRevE.94.043311} and recent advances in numerical techniques for lattice fermions \cite{Chandrasekharan:2009wc,Wang:2015rga,PhysRevB.93.155117,Huffman:2017swn,LiYao2019} are allowing us in particular to compute various critical exponents more accurately than before \cite{PhysRevLett.108.140404,PhysRevD.88.021701}. In some cases we have also discovered new and unexpected universality classes \cite{PhysRevD.93.081701,PhysRevD.96.034506} where it is believed that the exotic critical points may be described by non-Abelian gauge theories \cite{PhysRevX.8.011026,PhysRevB.97.125112}.

Despite the tremendous recent progress in the field, properties of even the simplest fermionic quantum critical points are very difficult to compute at the same level of accuracy as their bosonic counterparts \cite{Pelissetto:2000ek}. Focusing on Gross-Neveu models, the critical points are often characterized by the parameter $N_f$ (the number of four-component Dirac fermion flavors) and the symmetry breaking pattern (which are usually classified as either $\mathbb{Z}_2$ (Ising), $U(1)$ (XY), $SU(2)$ (Heisenberg)). In some studies the breaking of $SU(2)\times SU(2)$ symmetry has also been considered \cite{Gracey:1993pd,Christofi:2006zt}. For completeness, in the appendix we discuss the simplest three universality classes from a Hamiltonian perspective and compile some of the critical exponents obtained so far with $N_f=1,2$ in \cref{tab:GNexp}. As can be seen from the table, consistency between analytic results (using techniques like the $\epsilon$-expansions, large-$N$ expansions, functional RG and the bootstrap approach) and Quantum Monte Carlo (QMC) results (usings lattice formulations) is only visible for the $N_f=1$ chiral-Ising universality. Even among the QMC results, there is often a lot of spread. Due to limitations of convergence and difficulties to go to higher orders in the expansion, continuum calculations cannot easily be improved beyond what is currently available. Similarly, errors in Monte Carlo calculations arise due to the small lattice sizes used in the calculations. In fact most calculations have been performed on rather small lattice sizes, with total number of spatial lattice sites $N_s \approx 1000$ or lower. Very few calculations with $N_s \approx 2500$ exist \cite{sorella}. This should be compared to lattice calculations of $N_s > 10,000$ that are easily feasible at most bosonic critical points. Thus, an important area of research which motivates our study is alternative fermion Monte Carlo methods.

Traditional Monte Carlo methods that are often used to study relativistic four-fermion field theories can be classified into two types. The more popular method is the auxiliary field quantum Monte Carlo (AFQMC) technique, in which the time to complete a single sweep scales as ${N_s}^3\beta$, where $N_s$ is the spatial lattice volume and $\beta$ is the inverse temperature \cite{Bercx:2017pit}. The other method is the Hybrid Monte Carlo (HMC) method, which was developed originally for lattice QCD calculations in the Lagrangian formulation \cite{Duane:1987de,PhysRevLett.98.051601}. This method scales a lot better with system size when fermion masses are non-zero. Several studies have recently used this method to study fermionic critical behavior  \cite{PhysRevB.96.195408,PhysRevB.98.235129}. However, when extended to four-fermion models in the massless limit, it still does not seem to outperform the AFQMC method \cite{PhysRevB.97.085144}.

Alternatively, cluster QMC algorithms, which have long provided an efficient way to solve a variety of bosonic problems, were extended to fermions several years ago using the meron cluster concept \cite{Chandrasekharan:1999ys}. Recently this was extended to the notion of fermion bags \cite{Chandrasekharan:2009wc,Chandrasekharan:2013rpa}. This extension was then used to perform calculations at a fermionic quantum critical point with $N_s\geq 3600$ in both the Lagrangian formulation \cite{PhysRevD.93.081701} and the Hamiltonian formulation \cite{Huffman:2017swn}. Although the Hamiltonian fermion bag method also scales as ${N_s}^3\beta$ like the AFMC method, the prefactor turns out to be much smaller and hence allows us to explore much larger lattices than the traditional AFQMC method. Another feature of the algorithm is that it can be formulated even in the continuous-time limit within the Hamiltonian framework as was demonstrated in \cite{Huffman:2017swn}. However, since quantum critical behavior should not in principle depend on discrete-time errors, the continuous-time limit may not be necessary. In this work we thus explore if a discrete-time formulation of the Hamiltonian fermion bag approach can help to accelerate the fermion bag algorithm further. In doing so we give large lattice results for both continuous-time ($N_s=4096$) and discrete-time ($N_s=10,000$) formulations.

Our paper is organized as follows. In \cref{sec:HLFT} we first explain how one can construct a new type of Hamiltonian lattice field theory (HLFT) inspired by the fermion bag approach. We also explain the differences between our HLFT with the traditional Lagrangian lattice field theory (LLFT) that is often studied. In \cref{sec:model} we explicitly construct our HLFT for studying the Gross-Neveu chiral-Ising critical point with $N_f=1$. We explain the differences between the continuous-time and discrete-time models. In \cref{sec:fb} we describe how our new HLFT leads naturally to the notion of fermion bags and in \cref{sec:updalgo} we explain how one can use the fermion bag ideas to speed up the Monte Carlo updates. In \cref{sec:stab} we explain the stabilization procedures we have used during our calculations. Our results are then presented in \cref{sec:results} for both continuous-time and discrete-time fermion bag methods and \cref{sec:conc} contains our conclusions.

\section{Hamiltonian lattice field theory}
\label{sec:HLFT}

Relativistic four-fermion models are naturally formulated using the Lagrangian formulation in the continuum since we can explicitly construct them to be invariant under space-time rotations. A Lagrangian lattice field theory (LLFT) is then obtained as usual by discretizing the continuum Lagrangian on a space-time lattice which preserves a subgroup of this symmetry. Unfortunately, a naive discretization results in the well-known fermion doubling problem \cite{Nielsen:1980rz}, and to construct the theory without fermion doubling while preserving important chiral symmetries requires a more elaborate formulation using domain wall or overlap fermions \cite{Kaplan:1992bt,Narayanan:1993sk,Hands:2015dyp,Hands:2016foa}. While these formulations take care of the doubling issue, they are much more computationally intensive and studies that use them are likely to be limited to small lattice sizes in the near future. Another approach that has been recently explored is the use of SLAC fermions  \cite{PhysRevD.96.094504,Lang:2018csk}. Due to their non-locality, one can in principle formulate any number of two-component Dirac fermions while preserving all the symmetries \cite{PhysRevD.14.1627}. While it is well-known that they create many undesirable features in gauge theories \cite{Karsten:1980wd,Pelissetto:1987ad}, there seems to be some optimism in the community that they may give reliable results in four-fermion field theories. In our opinion this is far from clear and needs further research. A cheaper and reliable alternative to study at least a limited class of fermionic quantum critical points in LLFT is to use the staggered fermion approach on a three dimensional cubic lattice. However, in this approach we can only access even numbers ($N_f = 2,4,...$) of four-component Dirac fermions without using rooted staggered fermions. The rooting of staggered fermions involve assumptions that may not be valid for quantum critical behavior of interest here \cite{JANSEN20043,Creutz:2007yg,Creutz:2007rk,kronfeld2007lattice,sharpe2006rooted}. Recently, this approach was used to study critical points in $N_f=2$ models with both Ising and Chiral universality very accurately using the fermion bag algorithm \cite{PhysRevLett.108.140404,PhysRevD.88.021701}. Unfortunately, the staggered fermion approach also breaks important flavor and chiral symmetries and there is some worry that the universality classes may be affected due to such lattice artifacts.

Instead of an LLFT, one can also construct a space-time lattice field theory to study fermionic critical points starting with a lattice Hamiltonian. 
We refer to this as Hamiltonian lattice field theory (HLFT) to contrast it with LLFTs. By construction a HLFT is asymmetric between space and time and contains a parameter $\varepsilon$ which controls the temporal lattice spacing. Although relativistic invariance seems to have been lost it can be recovered if the quantum critical point is 
relativistic where the dynamical critical exponent $z=1$. For fermionic problems a Hamiltonian approach can in fact help in reducing the fermion doubling by a factor of two. Consider for example, free lattice staggered fermions hopping on a square lattice described by the Hamiltonian
\begin{align}
H = - t \sum_{x,\hat{d}} \sum_{a=1}^{N_f}\eta_{x,\hat{d}} (c^\dagger_{x,a} c_{x+\hat{d},a} + c^\dagger_{x+\hat{d},a} c_{x,a}),
\label{eq:sussferm}
\end{align}
where $t$ is an energy scale, $x$ labels spatial lattice sites on a square lattice and $\hat{d}$ a unit vector to the neighboring sites in the positive direction. The staggered fermion phase factors, $\eta_{x,\hat{d}}$, introduce a $\pi$-flux on each square plaquette \cite{PhysRevD.16.3031}. It is easy to argue that this lattice Hamiltonian describes $N_f$ four-component massless Dirac fermions in the continuum. Note that unlike LLFT, with HLFT we are no longer restricted to an even number of four-component Dirac flavors. However, the HLFT approach only preserves a discrete $\mathbb{Z}_2$ chiral symmetry.

Constructions of lattice field theories starting from Hamiltonians is not new and is often used in the condensed matter literature. In particular, all previous lattice Hamiltonian calculations using a discrete time path integral formulation should naturally be referred to as HLFTs based on our definition above. However, while one assumes $\varepsilon$ is small in such an approach, this may not be necessary to study properties of a quantum critical point. In our work we explore the possibility that we can perhaps even choose $\varepsilon=1$ and think of HLFTs as new types of lattice field theories with an asymmetry between space and time built into their formulations. In HLFT, physical lattice spacings in space and time are measured using physical scales as usual. One obvious worry is whether the universality class could change at large temporal lattice spacings since we are far from the Hamiltonian limit. An important result of our work is that this may not happen when $\varepsilon \approx 1$, suggesting that relativistic four-fermion field theories can indeed be formulated with a reduced fermion doubling using asymmetric space-time lattice formulations.

Another important aspect of our proposed HLFTs is that the choice of Hamiltonian is inspired by the idea of fermion bags. So in our approach to study $N_f$ free massless four-component Dirac fermions, instead of \cref{eq:sussferm} we replace it with
\begin{align}
H = - \sum_{x,\hat{d}} H_{x,\hat{d}},
\label{eq:fbferm}
\end{align}
which is a sum of nearest neighbor bond Hamiltonians, given by
\begin{align}
H_{x,\hat{d}} = \omega_{x,\hat{d}} e^{2\alpha_{x,\hat{d}} \sum_{a=1}^{N_f} (c^\dagger_{x,a} c_{x+\hat{d},a} + c^\dagger_{x+\hat{d},a} c_{x,a})}.
\label{eq:hlocal}
\end{align}
Clearly fermions described by \cref{eq:fbferm} are no longer free, but since all four-fermion interactions are perturbatively irrelevant in $2+1$ dimensions, it is still easy to argue that this modified Hamiltonian also describes $N_f$ free massless four-component Dirac fermions at long distances at sufficiently small values of $\alpha_{x,d}$. In order to match with \cref{eq:sussferm} we choose $2\omega_{x,d} \alpha_{x,d} = t\eta_{x,\hat{d}}$ and tune $\alpha_{x,d}$ to be small. We have recently shown that there are no sign problems as long as $\omega_{x,d} > 0$ \cite{Huffman:2015szi}. 

Our approach is not computationally elegant to study the free fermion critical point. On the other hand our goal is not to study that critical point, but rather some other interacting critical point connected to the free massless fermion phase. For such a study \cref{eq:fbferm} is a useful base Hamiltonian to begin with. For example, in the next section we will show how we can access an interesting fermionic critical points using \cref{eq:fbferm} by varying $\omega_{x,d}$ and $\alpha_{x,d}$ as a function of a coupling $V$. We can also introduce additional interactions and explore a richer phase diagram. For example we can add interactions of the form
\begin{align}
H_{\rm Hubb} = U \sum_x \Big(\ \sum_{a=1}^{N_f} \big(n_{x,a} - 1/2\big)\ \Big)^2,    
\end{align}
where $n_{x,a}=c^\dagger_{x,a}c_{x,a}$ is the fermion occupation number for a given flavor $a$, and continue to use fermion bag methods to understand the physics. 


\section{$N_f=1$ Ising Criticality}
\label{sec:model}

Our main goal in this section is to show how one can develop a fermion-bag inspired HLFT approach to study fermionic quantum critical points. In order to explain our approach, here we explicitly construct the HLFT for studying the chiral-Ising critical point with $N_f=1$ four-component Dirac fermions. The Hamiltonian we consider is the one we introduced in \cref{eq:fbferm} and \cref{eq:hlocal}, and for $H_{x,\hat{d}}$ we fix $N_f=1$, which gives us
\begin{align}
H_{x,\hat{d}} = \omega_{x,\hat{d}} e^{2\alpha_{x,\hat{d}} (c^\dagger_x c_{x+\hat{d}} + c^\dagger_{x+\hat{d}} c_x)}.
\label{eq:hlocal1f}
\end{align}
It is easy to verify that by choosing 
\begin{equation}
\begin{aligned}
\omega_{x,\hat{d}} &= \left(t^2/V\right)\left(1-\left(V/2t\right)^2\right),\\ \sinh 2\alpha_{x,\hat{d}}&= \left( V/t\right) / \left(1-\left(V/2t\right)^2\right) \eta_{x,d},\\ \cosh2\alpha_{x,\hat{d}} &= \left(1+\left(V/2t\right)^2\right) / \left(1-\left(V/2t\right)^2\right),
\end{aligned}
\end{equation}
our Hamiltonian is equivalent to 
\begin{equation}
\begin{aligned}
\small
H = & \sum_{x, \hat{d}}\left[ -t \eta_{x,\hat{d}} \left(c_x^\dagger c_{x+\hat{d}} + c^\dagger_{x+\hat{d}} c_x\right)\right. \\
&\left.\qquad+ V\left(n_x -\frac{1}{2}\right) \left(n_{x+\hat{d}}-\frac{1}{2}\right)\right].
\end{aligned}
\label{tvmodel}
\end{equation}
When $V$ is small, our model is in a massless fermion phase, while beyond some critical coupling $V_c$ the $\mathbb{Z}_2$ chiral symmetry in our model breaks spontaneously and fermions become massive. 

\begin{figure*}
\begin{center}
  \includegraphics[width=0.44\textwidth]{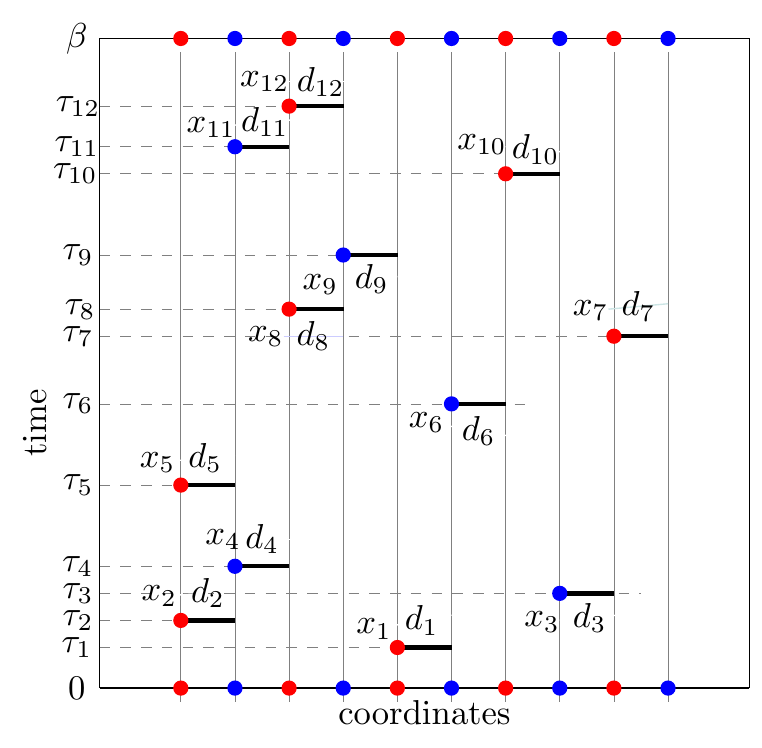}
   \includegraphics[width=0.48\textwidth]{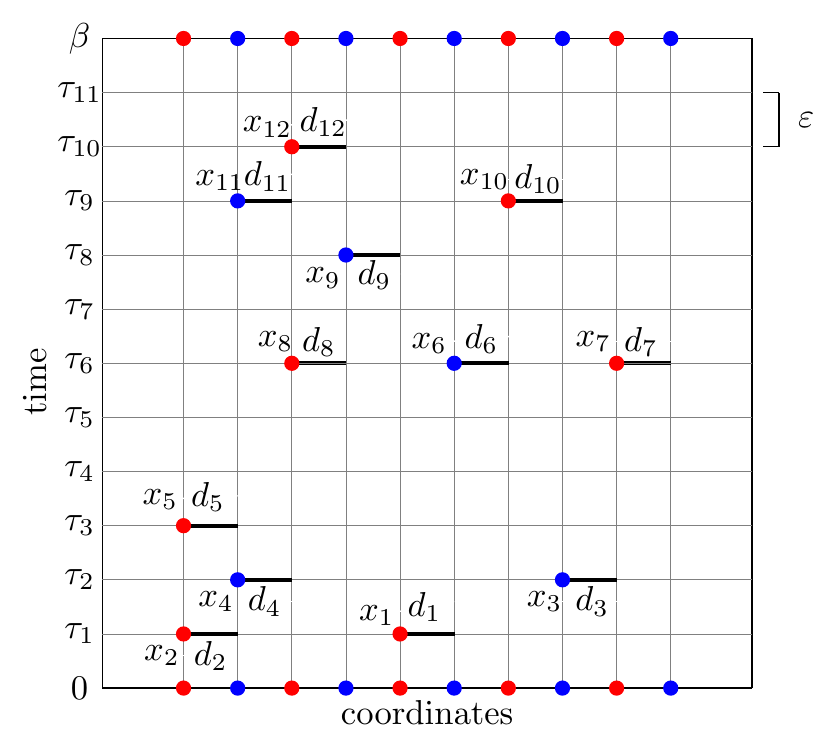}
  \end{center}
 \caption{Illustration of the continuous-time (left) and discrete-time (right) bond configurations $[\tau,x,\hat{d}]$ that define the respective partition functions. In both images, the spatial coordinates are colored red and blue to denote the bipartite nature of the lattice. \textit{Left}: The illustration shows a configuration with $k=12$. Spatial lattice site are shown on the horizontal axis, and the continuous imaginary time is shown on the vertical axis. Each bond contains two nearest neighbor spatial sites (but is defined by a site $x$ and a direction $\hat{d}$ in our notation) and is assigned a time coordinate $\tau$. \textit{Right}: The illustration shows a configuration again with $k=12$ but in the discrete-time HLFT. Here there are $N_t$ time slices in the model where bonds can exist. Importantly, multiple bonds can have the same time label $\tau$ so long as they have no sites in common.}\label{fig:confs}
\end{figure*}

We distinguish two types of HLFT partition functions depending on the temporal lattice spacing $\varepsilon$, one where $\varepsilon \neq 0$ (which we refer to as the discrete-time model) and the other where $\varepsilon \rightarrow 0$ (which we refer to as the continuous-time model). The partition function of the HLFT for the \textit{continuous-time} model, $Z=\mathrm{Tr}\big(e^{-\beta H}\big)$, can be constructed using the stochastic series expansion (SSE) approach \cite{PhysRevB.93.155117}:
\begin{equation}
\begin{aligned}
Z &= \ \sum_k\ \int \  [d\tau] 
\\
& \; \times\sum_{\left[\langle x,\hat{d}\rangle\right]}{\rm Tr}
\Big(H_{x_k,\hat{d}_k}(\tau_k) ... H_{x_2,\hat{d}_2}(\tau_2) H_{x_1,\hat{d}_1}(\tau_1)\Big),
\end{aligned}
\label{partition}
\end{equation}
where we define $H_{x_k,\hat{d}_k}(\tau_k) \equiv H_{x_k,\hat{d}_k}$ at time $\tau_k$ and 
there are k insertions of the bond Hamiltonian $H_{x,\hat{d}}$ inside the trace at times $\tau_1 \leq \tau_2 \leq ... \leq \tau_k$. The symbol $[d\tau]$ represents the k time-ordered integrals and the bond configuration $[\langle x, \hat{d}\rangle] = [\langle x_1, \hat{d}_1\rangle,\langle x_2, \hat{d}_2\rangle, ...\langle x_k, \hat{d}_k\rangle]$. The trace in \cref{partition} is evaluated in the fermionic Fock space. Note that a \textit{configuration} ${\cal C}$ in the "continuous-time" model is given by the set of $k$ bonds $[\tau,x,\hat{d}]$. One such configuration with $k=12$ is shown on the left in Figure \ref{fig:confs}. It is important to note that two bonds never appear at the same time in the continuous-time model and are always time ordered.

We can use \cref{partition} to construct the partition function of the HLFT for the discrete-time model with a temporal lattice spacing $\varepsilon \neq 0$ by replacing the integrals $[d\tau]$ by a sum. This then leads to the following expression for the discrete-time partition function:
\begin{equation}
\begin{aligned}
Z&=\ \sum'_{\left[\langle \tau,x,\hat{d}\rangle\right]} (\varepsilon)^k\\
& \qquad \times {\rm Tr}
\Big(H_{x_k,\hat{d}_k}(\tau_k) ... H_{x_2,\hat{d}_2}(\tau_2) H_{x_1,\hat{d}_1}(\tau_1)\Big).
\end{aligned}
\label{partitiondiscrete}
\end{equation}
Here $\varepsilon = \beta / N_t$, with $N_t$ being a finite number of time slices. We have also combined the sum over $k$, the integral $[d\tau]$ and the sum over bond configurations $[\langle x, \hat{d}\rangle]$ into a single sum over discrete time bond configurations $\left[\langle \tau, x,\hat{d}\rangle\right]$. These configurations are similar to the continuous time configurations, except that bonds can only exist on the allowed discrete time coordinates. However, we now allow several bonds to appear on the same time slice as long as they do not touch each other. This is consistent with the condition of time ordering $\tau_1 \leq \tau_2 \leq ... \leq \tau_k$, since the bond operators $H_{x,d}$ that do not touch each other commute with each other. Thus the order of the bonds within a timeslice has no effect on the weight. These additional constraints are denoted by the symbol ``$\ '\ $'' over the sum. A discrete time configuration is illustrated on the right in Figure \ref{fig:confs}. It is easy to argue that in the limit $\varepsilon \rightarrow 0$ the discrete-time partition function defined in \cref{partitiondiscrete} is equal to the continuous-time partition function defined in \cref{partition}, and we have numerically verified this fact as well.

We can now define the Boltzmann weight of every configuration of bonds in the both the continuous-time and discrete-time models through the expression
\begin{align}
\Omega([\langle \tau,& x,\hat{d}\rangle]) \ = \ \nonumber \\
& (\varepsilon)^k \ {\rm Tr}
\Big(H_{x_k,\hat{d}_k}(\tau_k) ... H_{x_2,\hat{d}_2}(\tau_2) H_{x_1,\hat{d}_1}(\tau_1)\Big).
\label{trweight}
\end{align}
It is clear from matching with \cref{partitiondiscrete} that when $\varepsilon$ and $N_t$ remain fixed, we get weights for the discrete-time model. On the other hand, in the limit $\varepsilon \rightarrow 0, N_t \rightarrow \infty$ so that $N_t \varepsilon = \beta$ remains fixed, we get the continuous-time model. In both cases, the fact that every $H_{x,\hat{d}}$ operator is an exponential of a fermionic bilinear operator is very useful to compute the trace in the Fock space. We can use the well-known BSS formula \cite{PhysRevD.24.2278}, which in our case yields the following determinant
\begin{align}
\Omega([\langle\tau,x,\hat{d}\rangle]) &\ = \  \Big(\prod_{x_k,\hat{d}_k} (\varepsilon \omega_{x_k,\hat{d}_k})\ \Big) \nonumber \\
\times &\det\left(\mathbbm{1} + h_{x_k,\hat{d}_k} ...h_{x_2,\hat{d}_2} h_{x_1,\hat{d}_1}\right)
\label{mcweight}
\end{align}
of an $N_s \times N_s$ matrix, where $N_s$ is the number of spatial lattice sites. In the above expression each $h_{x_i, \hat{d}_i}$ is an $N_s \times N_s$ matrix corresponding to the $H_{x_i,\hat{d}_i}$ operator in the one-particle basis. It is easy to verify that $h_{x_i, \hat{d}_i}$ are identity matrices except for a $2\times 2$ block, $\mathcal{H}_{x,\hat{d}}$, with entries located at the intersections of the rows and columns of the spatial sites that touch the bond $\langle x_i, \hat{d}_i \rangle$. The block is given by
\begin{align}
\mathcal{H}_{x,\hat{d}} = \left(\begin{array}{cc}
    \cosh 2\alpha_{x,\hat{d}} & \sinh 2\alpha_{x,\hat{d}}\\
    \sinh 2\alpha_{x,\hat{d}} & \cosh \alpha_{x,\hat{d}}
    \end{array}\right).
    \label{matblock}
\end{align}
We emphasize that the unusual form of the Hamiltonian in \cref{eq:hlocal} or \cref{eq:hlocal1f} was important to find a closed form expression for the weight $\Omega([\tau,x,\hat{d}])$ which can also be shown to be non-negative \cite{PhysRevB.89.111101,Li:2016gte}. Furthermore, as we will explain in the next section, the local nature of the bond operators also helps us construct efficient fermion bag algorithms. Note also that the expression for $\Omega([\langle\tau,x,\hat{d}\rangle])$ is the same whether it is in discrete time or continuous time as long as the bond configuration $[\langle\tau,x,\hat{d}\rangle]$ is the same. The quantum Monte Carlo algorithm then consists of updating a Markov chain of bond configurations $[\langle\tau,x,\hat{d}\rangle]$ by proposing changes to the current configuration and accepting/rejecting the proposed change using probabilities that satisfy detailed balance and are ergodic. We discuss the details of the algorithm for the continuous-time and discrete-time formulations in the next section.

The partition function of the discrete-time model \cref{partitiondiscrete} can also be expressed as a Grassmann integral of an unusual lattice field theory
\begin{align}
Z \ =\ \int \prod_{x,t}[d\overline\psi_{x,t} d\psi_{x,t}]\ e^{-S(\overline\psi,\psi)}
\end{align}
where we imagine that at every space-time lattice site $(x,t)$ there are two Grassmann values fields $\psi_{x,t},\overline{\psi}_{x,t}$, and the Euclidean action is given by
\begin{align}
& e^{-S(\overline\psi,\psi)}\ =\ e^{ \sum_{x,t} 
(\overline{\psi}_{x,t+1}-\overline{\psi}_{x,t})
\psi_{x,t}} \times \nonumber \\
& \prod_{x,t,\hat{d}}  \Big(  1  +  
 \varepsilon \omega_{x,\hat{d}} e^{(\overline{\psi}_{x,t+1},\overline{\psi}_{x+d,t+1})({\cal H}_{x,d}-\mathbf{1}_{x,d})
(\psi_{x,t},
\psi_{x+\hat{d},t})^T}\Big),
\end{align}
where we define $\mathbf{1}_{x,d}$ as a $2 \times 2$ unit matrix in the same space as ${\cal H}_{x,d}$. The Grassmann fields are assumed to satisfy anti-periodic boundary conditions in Euclidean time as usual. Note that it is difficult to identify the action $S(\overline\psi,\psi)$ in closed form for general values of $\varepsilon$, but when $\varepsilon$ becomes small (i.e., close to the continuum limit) we obtain 
\begin{align}
& S(\overline\psi,\psi)\ =\ -\sum_{x,t} 
(\overline{\psi}_{x,t+1}-\overline{\psi}_{x,t})
\psi_{x,t} \nonumber \\
& - \sum_{x,t,\hat{d}}\varepsilon \ \omega_{x,\hat{d}}\ 
e^{(\overline{\psi}_{x,t+1},\overline{\psi}_{x+d,t+1})({\cal H}_{x,d}-\mathbf{1}_{x,d})
\left(\psi_{x,t},
\psi_{x+\hat{d},t}\right)}.
\end{align}
The exponential terms in the action is related to the unusual form of the Hamiltonian. Note that the action is still local and asymmetric between space and time. 
 

\begin{figure*}
\begin{center}
\includegraphics[width=0.48\textwidth]{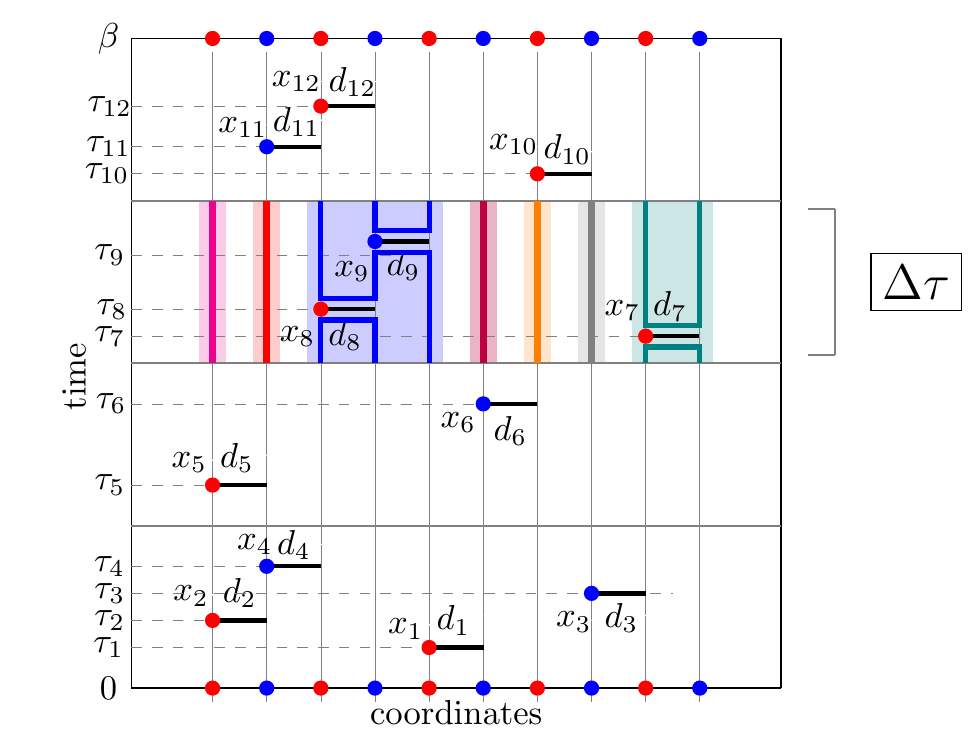}
 \includegraphics[width=0.465\textwidth]{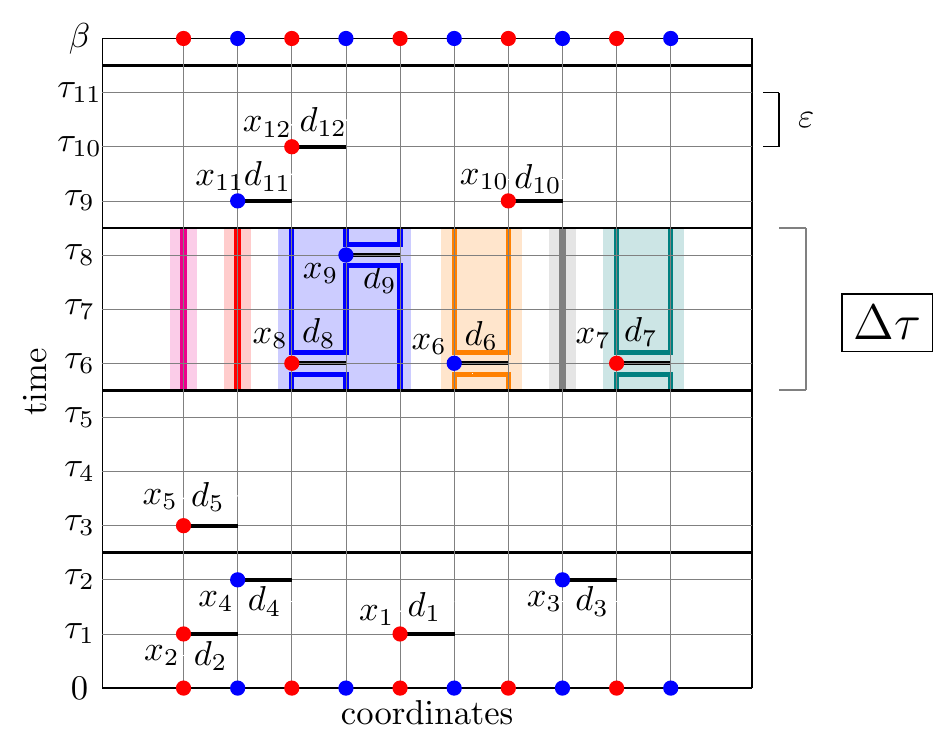}
 \end{center}
\caption{ Illustration of fermion bags within a time interval $\Delta \tau$ in the continuous-time model (left figure) and in the discrete-time model (right figure). There are seven fermion-bags in the continuous-time model and six fermion-bags in the discrete-time model. The time interval $\Delta \tau$ contains three time slices in the discrete-time model.
The matrix $\tilde{h}_{\Delta \tau}$ is the time ordered product of $h_{x,d}$ for the bonds that appear within $\Delta \tau$ in each case and hence is block diagonal within regions of fermion-bags.
\label{algorithms2}}
\end{figure*}

\section{Fermion Bags}
\label{sec:fb}

In addition to allowing us to compute the weight  $\Omega([\langle \tau,x,\hat{d}\rangle])$ easily, the local bond operators in \cref{eq:hlocal} or \cref{eq:hlocal1f} also help us in defining the notion of fermion bags in HLFT,  extending previous ideas discussed within the LLFT approach  \cite{PhysRevD.82.025007,Chandrasekharan:2013rpa}. Similar to the LLFT approach, definitions of fermion bags are not unique and the ideas we present here must be considered as one among other definitions that are possible. For example, it is possible to extend the notion of meron clusters (defined in \cite{Chandrasekharan:1999ys}) to the concept of fermion bags. In the definition we wish to explore here we first note 
that the bond operators commute with each other,
\begin{equation}
\left[H_{x,\hat{d}},H_{x',\hat{d}'}\right] = 0,
\label{comop}
\end{equation}
so long as the bonds $\langle x,\hat{d}\rangle$ and $\langle x', \hat{d}'\rangle$ do not share lattice sites. This also means that for the same bonds 
\begin{equation}
    \left[h_{x,\hat{d}},h_{x',\hat{d}'}\right] = 0.
\end{equation}
We can view $H_{x,\hat{d}}$ as creating entanglement between the two sites in the bond $\langle x,\hat{d}\rangle$, which means all spatial sites that are connected by bonds to each other (at various times) become entangled with each other. We can then study how spatial sites become entangled with each other as we focus on a fixed time interval $\Delta \tau$. Within this interval we define each such group of entangled spatial sites, which are connected due to bonds that are within that time interval, as a \textit{fermion bag}. Note that we consider each site that is not connected to any bonds (again, within the time interval) to form its own fermion bag. This definition of a fermion bag is dependent on the width of the time interval $\Delta \tau$.

In the extreme case, if we set the time interval to be the full extent of the imaginary time (i.e., $\Delta \tau = \beta$), typically all spatial sites will fall into a single fermion bag. This can be seen in Figure \ref{fig:confs}, where we see that in both the continuous-time and discrete-time configurations, all sites become entangled and thus belong to a single fermion bag. On the other hand, by decreasing $\Delta \tau$ we can reduce entanglement and increase the number of decoupled fermion bags. This is illustrated in \cref{algorithms2}. For the continuous-time model (left figure) we show an example where we have chosen $\Delta \tau = \beta / 4$. There we note that the spatial lattice splits into seven fermion bags. Similarly, in the discrete-time model (right figure) we show the case $\Delta \tau = 3\varepsilon$, where the time extent involves a bundle of three timeslices. In this case we note that there are six 
different fermion bags.

There are similarities and differences between the fermion bags we have defined above in the HLFT approach and those in the LLFT approach. In the LLFT approach the size of the coupling controls the size of the fermion bags, while in the HLFT definition above it is the temperature. Near the quantum critical point, it is more natural to expect that spatial entanglement decreases as temperature increases which is captured in the above definition. This feature can be used to construct fast updates even near quantum critical points. For example, one of the steps in the update process involves the calculation of the the $N_s\times N_s$ matrix $\tilde{h}_{\Delta \tau}$ which is the time ordered product of $h_{x,d}$ matrices corresponding to bonds that appear within the time region $\Delta \tau$. This matrix can be calculated very efficiently if space-time is split into several decoupled regions. Note that if we choose $\Delta \tau = \beta$ we get
\begin{align}
\tilde{h}_\beta = h_{x_k,\hat{d}_k} ...h_{x_2,\hat{d}_2} h_{x_1,\hat{d}_1}
\end{align}
which appears in \cref{mcweight}. We could imagine dividing the entire $\beta$ into several smaller Euclidean time regions and then computing $\tilde{h}_{\Delta \tau}$  efficiently for each region and then combining each of these results for computing $\tilde{h}_\beta$. We will devise such a strategy below, when we discuss the update algorithms.



In the discrete-time formulation we can speed up algorithms even further due to another concept of fermion bags defined in fixed background configurations \cite{PhysRevD.93.081701}. For example, consider the update of a single time slice in the discrete time formulation of the partition function. Since each bond change appears through the matrix $h_{x,d}$, which only affects two rows and two columns, it is possible to show that the ratio of the Boltzmann weights of a given background configuration and another configuration where $k$ bonds have been updated on a single time slice is given by--at most--a determinant of a $2k \times 2k$ matrix independent of the spatial size $N_s$. Thus, these $2k$ sites can be viewed as a fermion bag in a given background configuration. This additional fermion bag concept helps us build faster and more stable updates within a time slice in the discrete time formulation.


It is well known that measuring observables in theories with massless fermions is tricky since they can be singular. Configurations that contribute to the partition function may not be the same ones that contribute to the observable. The best known example of such a singular behavior is the chiral condensate in one flavor QCD with massless quarks. While gauge field configurations with a topological charge do not contribute to the partition function, the charge-one sector contributes to the chiral condensate. Thus an algorithm that samples the configurations of the partition function alone is not sufficient to measure the chiral condenstate. This type of singularity can also occur near fermionic critical points and hence we believe it is important to develop algorithms for each observable separately. Another issue is that it is important to make sure that the fermion bag concept remains applicable even in configurations that contribute to the observables. As we will see below, this is not difficult to achieve.

In this work we focus on computing the equal-time correlation function of the staggered mass order parameter through the operator
\begin{align}
C\ &=\ (-1)^{L/2} \left(n_0-1/2\right)\left(n_{L/2}-1/2\right)\ 
\label{corrfnop}
\end{align}
for the continuous-time limit and for the case with $\varepsilon=1$. Note that $(-1)^x (n_x-1/2)$ is the "staggered mass" operator, which acts as the order parameter for the chiral Ising transition in the $N_f=1$ Gross Neveu model. The correlation function is given by the expectation value
\begin{equation}
\left\langle C \right\rangle = \frac{1}{Z}\ {\rm Tr}\left(C e^{-\beta H}\right).
\label{corrfn}
\end{equation}
Following the steps of the previous section, we can write 
\begin{align}
{\rm Tr}\left(C e^{-\beta H}\right) \ & = 
\sum'_{\left[\langle \tau,x,\hat{d}\rangle\right]} (\varepsilon)^k \mathrm{Tr}\Big(H_{x_k,\hat{d}_k}(\tau_k) ...
\nonumber \\
& ... C\left(\tau_C\right) ...
H_{x_2,\hat{d}_2}(\tau_2) H_{x_1,\hat{d}_1}(\tau_1)\Big),
\end{align}
where we have introduced the operator $C$ at the Euclidean time $\tau_C$. Due to the cyclic property of the trace, the operator $C$ could be located anywhere in imaginary time. However, for individual configurations $[\langle\tau,x,\hat{d}\rangle]$, as defined in Section III, the operator $C$ will cause a configuration to have a different weight depending on where it is placed in imaginary time, because it does not commute with the other insertions of $H_{x,\hat{d}}$. Hence, in our algorithm we enlarge our configuration space by sampling two types of configurations: $[\langle\tau,x,\hat{d}\rangle]$ which contribute to the partition function and $[\langle\tau,x,\hat{d}\rangle,\tau_C]$ which defines the same configuration with an additional imaginary time location of the $C$ operator \footnote{If bond operators appear at the same time slice along with the operator $C$ we assume the operator $C$ appears before the bond operators}. We can combine the two types of configurations by assuming that in both cases we introduce a new operator $C_n$ at the time $\tau_C$ such that $C_{n=0} = 1$ and $C_{n=1} = C$. Thus the configuration space of our algorithm is always labeled using $[\langle\tau,x,\hat{d}\rangle,\tau_C]$, and we define corresponding weights $\Omega_n$ for the two types of configurations as
\begin{equation}
\begin{aligned}
\Omega_n&([\langle\tau ,x,\hat{d}\rangle,\tau_C])= \left(\varepsilon\right)^k  {\rm Tr}\left(H_{x_k,\hat{d}_k}\left(\tau_k\right) ...\right.\\
&\qquad \qquad\left. ... C_n \left(\tau_C\right)  ... H_{x_2,\hat{d_2}}\left(\tau_2\right) H_{x_1,\hat{d_1}}\left(\tau_1\right)\right).
\label{omegaweights}
\end{aligned}
\end{equation}

Similar to \cref{mcweight}, we can again use the BSS formula to compute $\Omega_n([\langle\tau ,x,\hat{d}\rangle,\tau_C])$, so long as $C$ is also constructed out of exponential of fermionic bilinear operators. Fortunately we can use $(n_x-1/2) = -e^{i\pi n_x}/2$ to construct $C$. With this choice it is easy to see that 
\begin{equation}
\begin{aligned}
&\Omega_n ([\langle\tau,x,\hat{d}\rangle,\tau_C]) =  \\
& \qquad\frac{1}{4}\prod_{x_k,\hat{d}_k}( \varepsilon \omega_{x_k,\hat{d}_k})\det\left(\mathbbm{1} + h_{x_k,\hat{d}_k}\left(\tau_k\right) \right. \\
&\left. \qquad \qquad\qquad ...c_n\left(\tau_C\right)...h_{x_2,\hat{d}_2}\left(\tau_2\right) h_{x_1,\hat{d}_1}\left(\tau_1\right)\right),
\end{aligned}
\end{equation}
where in addition to $h_{x_i, \hat{d}_i}$, which we already encountered in \cref{mcweight}, we have introduced the $N_s \times N_s$ matrix $c_n$ which is the unit matrix for $n=0$ and the diagonal with $+1$ at all spatial sites except at $x=0$ and $x=L/2$ where it is $-1$. Using these definitions for the weights $\Omega_n ([\langle\tau,x,\hat{d}\rangle,\tau_C])$, the correlation function is given by
\begin{equation}
\left\langle C \right\rangle = \frac{\sum_{[\langle\tau, x,\hat{d}\rangle,\tau_C]} \Omega_1 ([\langle\tau ,x,\hat{d}\rangle,\tau_C])}{\sum_{[\langle\tau', x',\hat{d}'\rangle,\tau_C']} \Omega_0 ([\langle\tau' ,x',\hat{d}'\rangle,\tau_C'])}.
\end{equation}
It is easy to note that the fermion bag concepts we introduced above for the $n=0$ sector are also valid for the $n=1$ sector. Thus allows us to sample both $n=0$ and $1$ sectors. 

Unfortunately, the weights $\Omega_n([\langle\tau,x,\hat{d}\rangle],\tau_C)$ are such that it is not easy for the algorithm to tunnel between the two sectors. In order to alleviate this problem we use a factor $f>0$ to reweight the $n=1$ sector. Still, the observable $\langle C\rangle$ suffers from large autocorrelation times since the ratio 
$\Omega_1([\langle\tau,x,\hat{d}\rangle],\tau_C)/\Omega_0([\langle\tau,x,\hat{d}\rangle],\tau_C)$ can fluctuate a lot. For this reason, instead of $\langle C\rangle$ we measure the ratio 
\begin{equation}
    \mathcal{N} = \frac{\Omega_1([\langle \tau,x,\hat{d},\rangle,\tau_C])}{\Omega_0([\langle \tau,x,\hat{d}\rangle,\tau_C]) + f \Omega_1([\langle \tau, x,\hat{d}\rangle,\tau_C])}.
\end{equation}
for each configuration $([\langle\tau' ,x',\hat{d}'\rangle],\tau_C')$ that we generate irrespective of the sector we are in. Taking a usual Monte Carlo average then allows us to compute $\langle \mathcal{N} \rangle$, which is given by the expression
\begin{equation}
\begin{aligned}
     \langle\mathcal{N}\rangle & = \sum_{[\langle\tau, x,\hat{d}\rangle,\tau_C]}\Omega_1([\langle \tau,x,\hat{d},\rangle,\tau_C]) \Big/ \\
     &\sum_{[\langle\tau', x',\hat{d}'\rangle,\tau_C']} \Big(\Omega_0([\langle \tau',x',\hat{d}'\rangle,\tau_C']) \\
     &\qquad\qquad\qquad\qquad + \ f \ \Omega_1([\langle \tau', x',\hat{d}'\rangle,\tau_C'])\Big).
     \end{aligned}
\end{equation}
We then find the observable using the relation 
\begin{align}
\left\langle C\right\rangle = \left\langle \mathcal{N}\right\rangle / (1-f\left\langle \mathcal{N}\right\rangle).    
\end{align} 
This eliminates the large fluctuations and autocorrelations in $\langle C\rangle$. Unfortunately, this also limits us since we are only able to compute one observable at a time, making additional observables expensive to compute.



\section{Update Algorithms}
\label{sec:updalgo}
In this section we will discuss the update algorithms in detail, distinguishing between the updates for the continuous-time and the discrete-time models. The former was introduced earlier in Ref. \cite{Huffman:2016cgh,Huffman:2017swn}. Algorithms for both models consist of four types of updates:
\begin{enumerate}
\item The \textit{sector-update} flips the sector from $n$ to $1-n$. This means the configuration with the operator $C_n$ inserted at time $\tau_C$ is replaced by the operator $C_{1-n}$. This is accomplished through a Metropolis accept/reject algorithm.
\item The \textit{bond-update} changes the bond configuration from $[\langle \tau, x,\hat{d} \rangle,\tau_C]$ to $[\langle \tau',x',\hat{d}'\rangle,\tau_C]$ without a change in the operator insertion at $\tau_C$. In the continuous-time model we divide $\beta$ into several smaller temporal regions $\Delta \tau$. We then sequentially pick every temporal region and perform bond updates within that region. In the discrete-time model we go through every time slice in a sequence and update the bonds on that time slice. Again the updates are performed using a local Metropolis accept/reject step within the chosen time region or the time slice.
\item The \textit{time-update} changes the imaginary time location $\tau_C$ where the operator $C_n$ is introduced. All possible time values for $\tau_C$ are chosen with equal probability. To make the update easy, this update is only performed in the $n=0$ sector, so that the new $\tau_C$ is always accepted.
\item The \textit{move-update} updates the temporal location of a bond chosen at random with the constraint that it never crosses another bond that shares one of its sites. Such an update does not change the weight of the configuration and is always accepted. For example, in the continuous-time model configuration shown in \cref{fig:confs}, the bond with coordinates $x_6,\hat{d}_6$ at $\tau_6$ can be moved to any time between $\tau_1$ and $\tau_{10}$ 
without changing the configuration weight. 
\end{enumerate}
We define a sweep as accomplishing a fixed number of each one of these four types of updates. Among these updates, the most time-intensive updates are the sector-update and the bond-update, as they require us to compute ratios of determinants that appear in the expressions for  $\Omega_n([\langle\tau,x,\hat{d}\rangle,\tau_C])$. Since these are similar to the weights one encounters in the AFQMC methods we will borrow some of the ideas from there. However, for efficient computation we will combine them with techniques that use the fermion bag concept. For example, an important distinction between the fermion bag algorithms on one hand, and the discrete-time auxiliary field formulations on the other, is that we can update a configuration over a $\Delta \tau$ that is significantly larger than the single time slice that are usually updated in AFQMC. This leads to the ability for more freedom in terms of the local updates since they no longer have to be completely sequential, and thus improvement in autocorrelation times, however larger $\Delta \tau$ can also lead to more stabilization difficulties in updating $G_B$ and we will show how fermion bags can be helpful for dealing with this problem as well. The time-update and move-update on the other hand are easy to implement and do not require time-intensive calculations.

\begin{figure*}
\begin{center}
\includegraphics[width=0.47\textwidth]{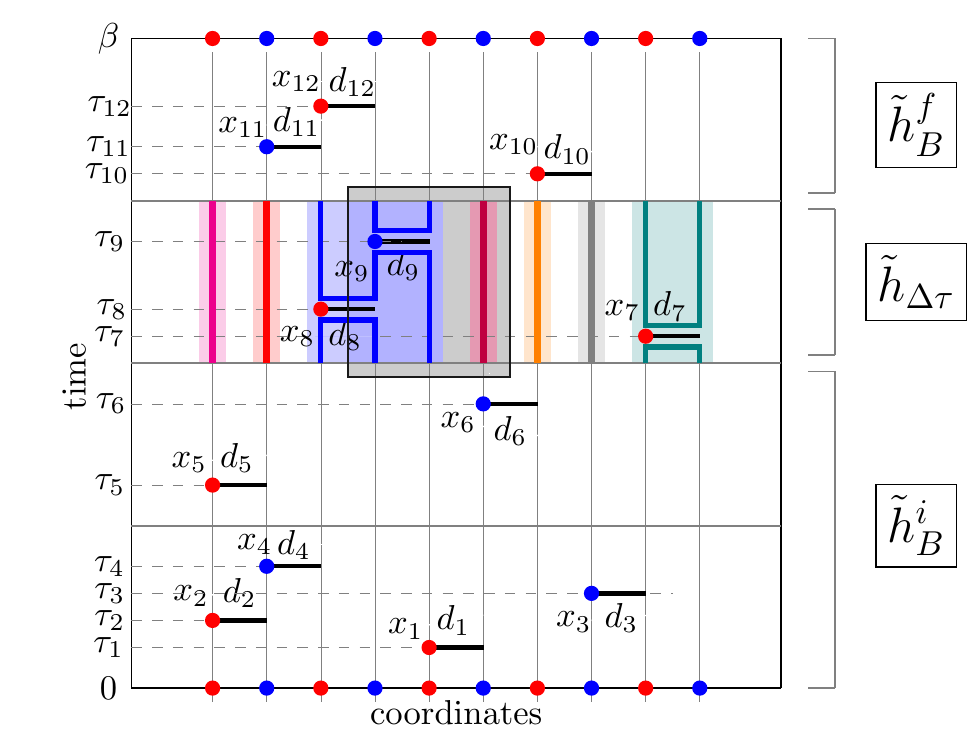}
\includegraphics[width=0.45\textwidth]{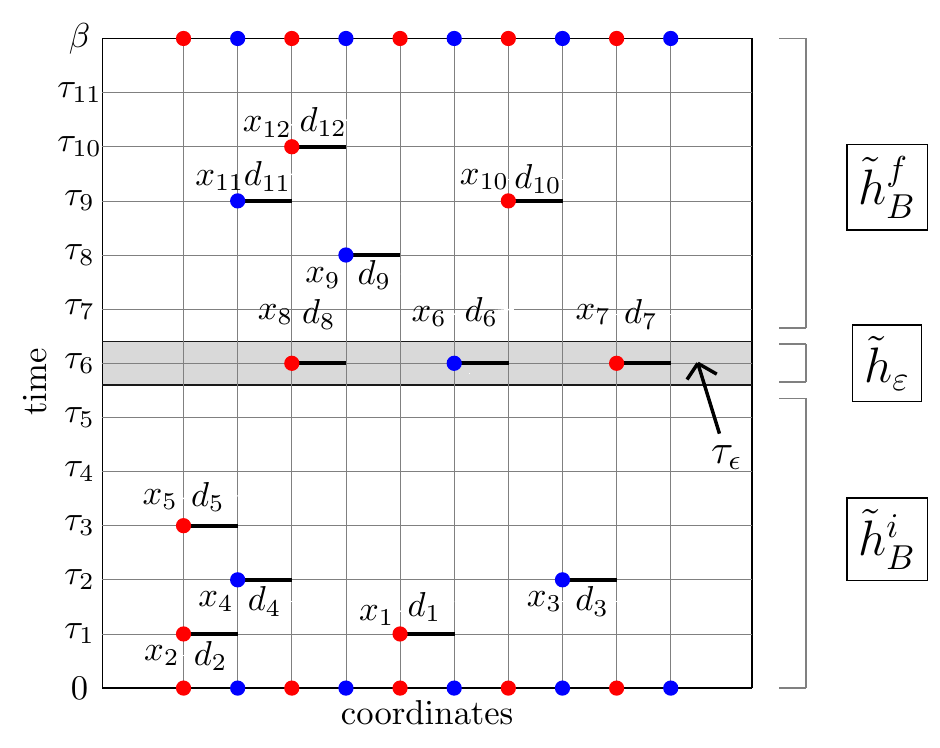}
\end{center}
\caption{Illustration of the bond update procedure in the continuous-time model (left) and the discrete-time model (right). A spatial slice with some width in time is first chosen where the bonds are updated. In continuous-time the width is $\Delta \tau$ obtained by dividing $\beta$ into several equal parts (left figure, where $\beta$ is divided into four parts). While in principle the same procedure is also possible in discrete-time, we can also chose a single time slice (right figure). Next, the background matrix $\tilde{h}_B = \tilde{h}_B^i \tilde{h}_B^f$ is computed. Bonds are updated in a spatial block within the chosen temporal region (shown as a grey box). In continuous-time this is a finite spatial region (left figure), while in discrete-time it is chosen to be the entire spatial lattice (right figure).
Using concepts of fermion bags fast updates are designed within each region.}\label{algorithms3}
\end{figure*}

Since both sector-updates and bond-updates use the Metropolis accept/reject algorithm, we need to compute the weight ratios $R$ between the final and initial configurations. For example
\begin{equation}
R = \frac{ \Omega_n([\langle \tau', x',\hat{d}'\rangle,\tau_C])}{\Omega_n([\langle \tau, x,\hat{d} \rangle,\tau_C])},
\label{ratio}
\end{equation}
is necessary to compute the transition probabilities during bond-updates. Here we assume that the final configuration obtained after the update, $[\langle \tau', x',\hat{d}'\rangle,\tau_C]$, and the initial configuration, $[\langle \tau, x,\hat{d}\rangle,\tau_C]$, differ by bonds only within a time region $\Delta \tau$. For the sector update the weight ratio is mathematically similar except that the numerator will instead be $\Omega_{1-n}([\langle \tau,x,d\rangle \rangle,\tau_C])$. We now discuss how we compute $R$ for the bond-update. The procedure for the sector-update is a straightforward generalization of our discussion here.

Let us assume that the region $\Delta \tau$ where the bonds are being updated lie after the bond at some initial time $\tau_i$, but before the bond at some final time $\tau_f$. Assuming the final bond is at some time $t_k$, we can define
\begin{align}
&\tilde{h}_B^i= h_{x_{i},\hat{d}_{i}}(\tau_{i}) h_{x_{i-2},\hat{d}_{i-2}}(\tau_{i-2})...h_{x_{1},\hat{d}_{1}}(\tau_{1})
\nonumber \\
&\tilde{h}_B^f = h_{x_{k},\hat{d}_{k}}(\tau_{k}) h_{x_{k-1},\hat{d}_{k-1}}(\tau_{k-1})...h_{x_{f},\hat{d}_{f}}(\tau_{f}),
\end{align}
for both initial and final configurations since the bonds outside the $\Delta \tau$ region do not change. Within the region where the bonds do change we define
\begin{align}
&\tilde{h}_{\Delta\tau} = h_{x_{f-1},\hat{d}_{f-1}}(\tau_{f-1}) ...h_{x_{i+1},\hat{d}_{i+1}}(\tau_{i+1})    
\end{align}
for the configuration $[\langle \tau, x,\hat{d}\rangle,\tau_C]$ and
\begin{align}
&\tilde{h}'_{\Delta\tau} = h_{x'_{f-1},\hat{d}'_{f-1}}(\tau'_{f-1}) ...h'_{x_{i+1},\hat{d}'_{i+1}}(\tau'_{i+1})    
\end{align}
for the configuration $[\langle \tau', x',\hat{d}'\rangle,\tau_C]$. It is then easy to verify that 
\begin{align}
R &= 
\frac{\det(\mathbbm{1} + \tilde{h}^f_B \tilde{h}'_{\Delta\tau} \tilde{h}^i_B)}
{\det(\mathbbm{1} + \tilde{h}^f_B \tilde{h}_{\Delta\tau} \tilde{h}^i_B)}
\ =\ 
\frac{\det(\mathbbm{1} + \tilde{h}_B \tilde{h}'_{\Delta\tau})}
{\det(\mathbbm{1} + \tilde{h}_B \tilde{h}_{\Delta\tau})}
\label{eq:ratio1}
\end{align}
where in the last step we have defined a single background matrix $\tilde{h}_B = \tilde{h}_B^i \tilde{h}_B^f$ which clearly does not change during the update of bonds within the time slice $\Delta \tau$.

As in AFQMC methods, in order to efficiently obtain the ratio $R$, the key quantity to compute is the $N_s \times N_s$ matrix (often referred to as the Green's function), which in our case is given by 
\begin{align}
G_B & = \left(\mathbbm{1} + \tilde{h}_B \tilde{h}_{\Delta \tau}\right)^{-1} \tilde{h}_B \tilde{h}_{\Delta \tau} \\
&= \mathbbm{1} -\left( \mathbbm{1} +\tilde{h}_B \tilde{h}_{\Delta \tau}) \right)^{-1},
\label{etgreens}
\end{align} 
This Green's function $G_B$ can be used for fast updates of bonds within the interval $\Delta\tau$ since we can express the weight ratio as
\begin{align}
R &= \det\left(\mathbbm{1} + G_B\ \Delta\right),\quad
\Delta = (h^{-1}_{\Delta \tau} h'_{\Delta \tau} - \mathbbm{1})
\label{eq:fastR}
\end{align}
The advantages of our fermion-bag inspired HLFT becomes clear at this step. Notice that the matrix $\Delta$ in \cref{eq:fastR} is zero everywhere except within a region obtained by the union of all fermion-bags touched by the updated bonds. We define the size of this spatial region as $s$, which implies that the matrix $\Delta$ is non-zero only within this $s\times s$ block. This means the computation of $R$ is reduced to the determinant of an $s\times s$ matrix defined as
\begin{align}
R &= \det\left(\mathbbm{1}_{s\times s} + (G_B)_{s\times s} (\Delta)_{s\times s}\right),
\label{continuousdif}
\end{align}
where $(G_B)_{s\times s}$ and $\Delta_{s\times s}$ are matrices restricted to the $s \times s$ block discussed above. Note that $s$ is obtained adding the sites corresponding to the updated bonds \textit{plus} the sites belonging to fermion bags that touch those bonds. So when the updates begin $s=0$ and as more and more number of bonds are updated $s$ begins to grow and the calculation of $R$ becomes more expensive. However, at no time does the entire spatial size enter the computation since all calculations are restricted to aN $s\times s$ block. 

\begin{figure*}
\begin{center}
  \includegraphics[width=0.48\textwidth]{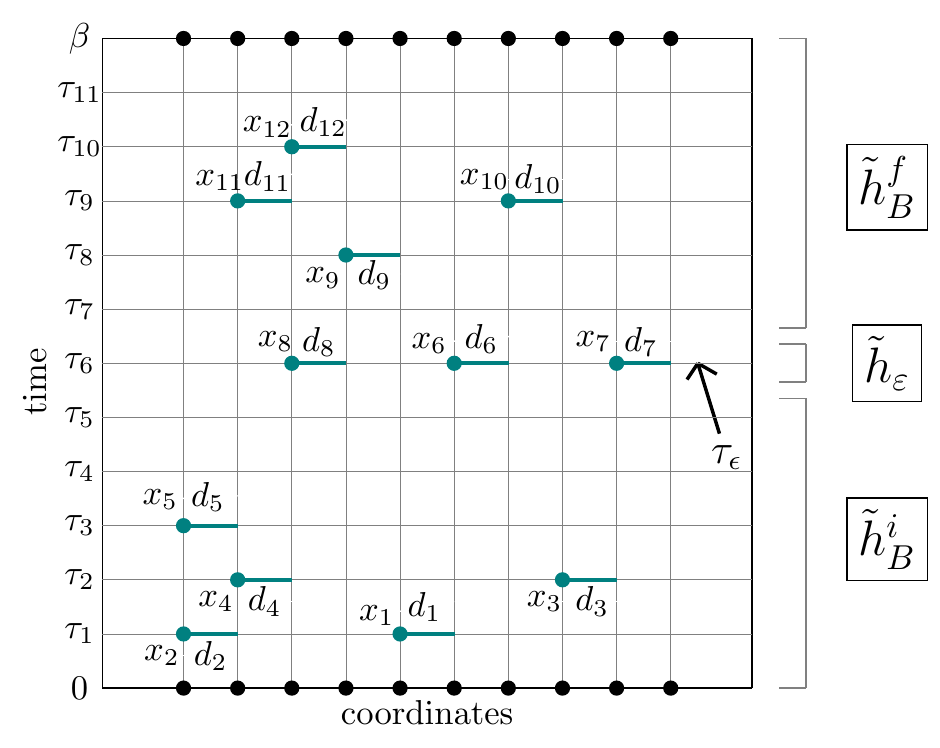}
  \includegraphics[width=0.48\textwidth]{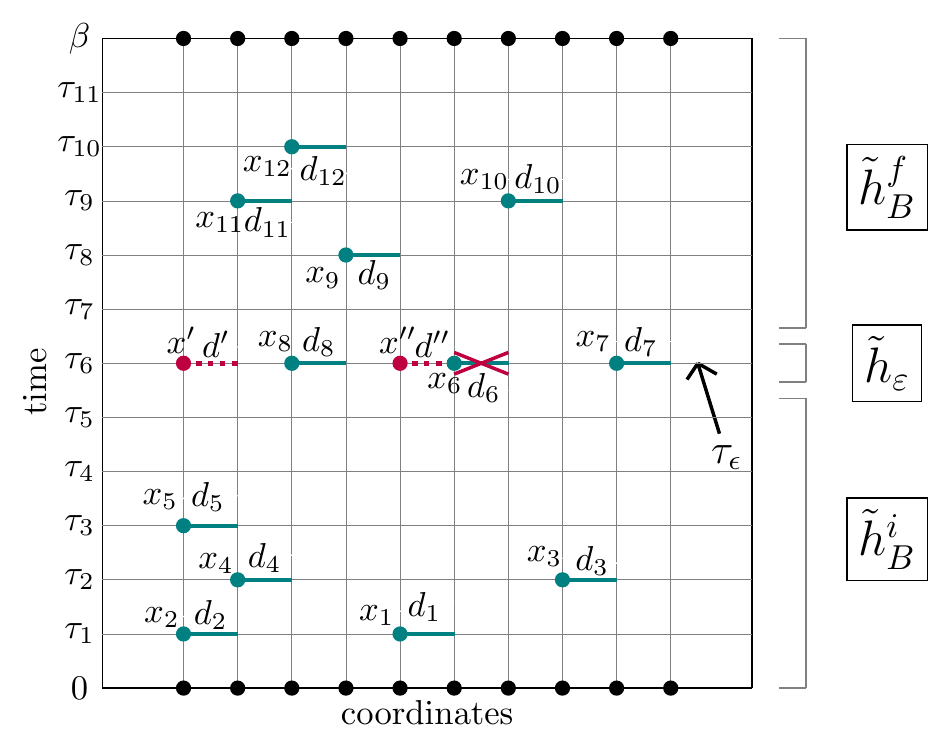}
  \end{center}
 \caption{An illustration of a bond-update in the discrete-time model on one of the chosen discrete time-slices. The configuration shown in the left is the initial configuration and that on the right is the final configuration. The bond labeled $(x_7,d_7)$ has been removed, and the two bonds labeled $(x',d'),(x'',d'')$ have been added. Since five sites are affected by the update $s=5$ in this example and the weight ratio $R$ can be written as the determinant of a $5\times 5$ matrix.}\label{discreteupdate}
\end{figure*}

In the continuous-time model, the temporal region $\Delta \tau$ is updated by choosing a random spatial block of sites at a time, as illustrated by the shaded gray box in the left side of  \cref{algorithms3}. This limits the sites affected during the update. Of course we choose several random blocks to update the entire spatial lattice. The size of each block is chosen to be on the order of the average fermion bag size. During the block update $s$ is chosen to be equal to the number of sites in the block \textit{plus} any sites outside of the block but are part of fermion bags that are partially inside the block. We call this larger set of sites a \textit{superbag} and $s$ is its size. 
Thus, during the Metropolis accept/reject step, rather than computing the determinant of a full $N_s \times N_s$ matrix, we only have to compute the determinant of an $s\times s$ matrix in \cref{continuousdif}. Although $s$ is usually small, the computation $\Delta$ can still suffer from stabilization issues which is well known in AFQMC. One cannot naively compute it as a product of the bond matrices $h_{x,d}$ that are being updated. This stabilization problem must be handled carefully, especially as $\Delta \tau$ intervals get larger. The problem is closely related to the physics of the model. Fortunately, for the model we are considering here, we can choose $\Delta \tau \approx 1/4$ and are able to deal with stabilization issues near the critical point as discussed in the next section.

In contrast to the continuous-time model, the discrete-time model with $\varepsilon = 1$ offers new advantages. We can update each time slice at a time since it is equivalent to choosing $\Delta \tau = 1$, which is much larger than what is feasible in the continuous-time model due to stabilization problems. The discrete time model has no stabilization problem during the computation of $\Delta$ since bonds do not touch on a single time-slice. To see this note that the weight ratio (\cref{eq:ratio1} and \cref{eq:fastR}) are replaced by
\begin{align}
R &= \frac{\det(\mathbbm{1} + \tilde{h}_B {\tilde{h}_\varepsilon}')}{\det(\mathbbm{1} + \tilde{h}_B {\tilde{h}_\varepsilon})} = \det(\mathbbm{1} + G_B \Delta).
\label{discretedif}
\end{align}
for the update of a single time-slice located at $\tau_\varepsilon$ (see right figure of \cref{discreteupdate}). Here $G_B (\tau_T) = (\mathbbm{1} + \tilde{h}_B \tilde{h}_{\tau_\varepsilon})^{-1} \tilde{h}_B \tilde{h}_{\tau_\varepsilon}$ where $\tilde{h}_B = \tilde{h}^i_B \tilde{h}^f_B$ is similar to the matrix defined earlier and is given by product of all the bond matrices $\tilde{h}_{x,\hat{d}}\left(\tau\right)$ that are not on the discrete time-slice $\tau_\varepsilon$, 
and $\tilde{h}_{\tau_\varepsilon}$ is the product of matrices that are in that time-slice. Correspondingly, we have $\Delta = (\tilde{h}_\varepsilon^{-1} \tilde{h}'_\varepsilon -\mathbbm{1})$, which is again zero everywhere except for in an $s\times s$ block, where $s$ now is the number of sites where the bonds have changed (either due to new a bond added or an old bond removed) as compared to the initial configuration. \cref{discreteupdate} gives an illustration of a time slice update in the discrete-time model. In this illustration two bonds have been added and one bond has been subtracted from the active time-slice. Because five sites are affected in total, $s=5$ and so the determinant in (\ref{discretedif}) will be of a $5\times 5$ matrix. Computing $h_\varepsilon$ or $h'_\varepsilon$ leads to no stabilization problems since bonds do not touch each other. Note that in contrast to the continuous-time version, the dimension of the matrix will grow as updates are made rather than remaining constant. This is similar to fermion-bag algorithms used in the Lagrangian picture \cite{PhysRevD.91.065035,PhysRevD.93.081701,Ayyar:2017xmi,PhysRevD.96.114506}, and means that instead of limiting ourselves to one limited spatial block at a time, we can update the background after $s$ gets past a maximum size that we set.


\section{Stabilization of Matrix Multiplication}
\label{sec:stab}
As is well-known from traditional AFQMC methods \cite{LOH1992177,Assaad2008}, the multiplication of many $N_s \times N_s$ matrices that may become necessary in the calculation of $R$ can suffer from stabilization issues if performed naively. There are three main numerical instabilities that we have to deal with: (1) computing $G_B$ from scratch and then updating it as we move on to subsequent time-intervals $\Delta \tau$ or time-slices $\tau_\varepsilon$, (2) updating $G_B$ as we change the spatial block or refresh the background within the same time-interval or time-slice, and (3) updating $\Delta$ after adding or removing bonds. We discuss our strategy for dealing with these three types of problems below.

Let us first consider the problem of computing $G_B$ using \cref{etgreens}, which involves computing $\tilde{h}_B$ and $\tilde{h}_{\Delta \tau}$ from the matrices $h_{x,\hat{d}}(\tau)$ and then performing further matrix operations such as multiplications, addition of the identity matrix, and computing inverses. Unfortunately, many of these operations cannot be done using straightforward matrix routines on a computer. The calculation may involve numbers at different scales that can be orders of magnitude apart, and the physics of the small scales can be completely lost in the process. In a typical auxiliary field Monte Carlo method, the operations are accomplished using the singular value decomposition (SVD) or Gram-Schmidt decomposition of the individual matrices, and operations are constructed carefully so that the information about small scales is not lost. This is time-consuming and it would be helpful to avoid it as much as possible. In our case, since each $\tilde{h}_{x,\hat{d}}$ matrix is non-trivial only in a $2\times 2$ block, we can indeed multiply several such matrices at a time without worrying about stabilization issues. Still, since the $2\times 2$ blocks contain exponential functions of the form $\cosh(2\alpha_{x,\hat{d}})$ and $\sinh(2\alpha_{x,\hat{d}})$, multiplying more than $5-10$ blocks can potentially lead to stabilization problems. The choice of a \textit{bundling} parameter, $n_b$, which is an integer that gives how many blocks can be combined stably at one time, is dictated by this problem and depends on the parameters $\alpha_{x,\hat{d}}$, with larger values requiring a smaller $n_b$ size.

Once $n_b$ has been set,  we think through the formation of $G_B$, which will involves a multiplication of all the $\tilde{h}_{\Delta \tau}$'s together, which in turn may involve the multiplication of several \textit{bundles}-worth of matrices $h_{x,\hat{d}}\left(\tau\right)$ (a new bundle is started after the naive multiplication of $n_b$ matrices that correspond to bonds that touch the same site.) These steps will need some kind of stabilization. Fortunately, we can accomplish this without using SVDs, as discussed below. First note that given a generic matrix $h_i$ whose matrix elements do not involve a large disparity of scales, the Greens function matrix $G_i = \left(\mathbbm{1}+h_i\right)^{-1} h_i = \mathbbm{1}-\left(\mathbbm{1} + h_i\right)^{-1}$ can be constructed without stability problems. Further $G_i$'s are very well behaved matrices and operations involving them are quite stable without the need of SVDs. Thus, given two generic matrices, $h_1$ and $h_2$, we can construct the Greens function for their produce $h_1 h_2$
using the identity
\begin{equation}
\begin{aligned}
    \left(\mathbbm{1}\right. &\left.+h_1 h_2\right)^{-1} = \left(\mathbbm{1}-G_2\right) \times \\
    & \big( \left(\mathbbm{1}-G_1\right)\left(\mathbbm{1}-G_2\right) + G_1 G_2 \big)^{-1}\left(\mathbbm{1}-G_1\right).
    \end{aligned}
    \label{buildg}
\end{equation}
As noted above the right hand side of \cref{buildg} does not suffer from stabilization issues. Thus, we can sequentially build $G_B$ from partial versions labeled as $G_{\Delta \tau} = (1+\tilde{h}_{\Delta \tau}) \tilde{h}_{\Delta \tau}$ associated with each time region $\Delta \tau$, which were in turn built from one or more different $G_{\Delta \tau, n_b}$ pieces corresponding to the different matrix bundles within $\Delta \tau$. As will be described at the end of the section, we can further split up these matrix combinations into smaller matrix operations using the fermion bag concept.
We can also use a similar procedure to update $G_B$ as we move to a different time slice as long as the regions $\Delta \tau$ or $\tau_\varepsilon$ are chosen sequentially.

Let us now consider the type-3 instability that occurs while computing ${\tilde{h}_{\Delta \tau}'}$ which enters the block matrix $\Delta$ in the continuous-time formula (\cref{continuousdif}). Here we assume $G_B$ has been calculated and stored, and we need to compute $\Delta$ for each accept/reject proposal. Instead of computing $\Delta$ and then computing $R$ using \cref{continuousdif} or \cref{discretedif}, we actually compute
the $s\times s$ matrix $F = [G_B]_{s\times s} [\tilde{h}_{\Delta \tau}^{-1}]_{s\times s}[\tilde{h}'_{\Delta \tau}]_{s\times s}$, and then calculate the ratio using the formula
\begin{equation}
\begin{aligned}
R &= \det([\mathbbm{1}-G_B]_{s\times s} + F) \\
&= |\det([\mathbbm{1}-G_B]_{s\times s} \mathcal{Q}^T + \mathcal{R})|
\label{detstab}
\end{aligned}
\end{equation}
where we are using the $RQ$ factorization of matrix $F$ into an upper triangular matrix $\mathcal{R}$ and an orthogonal matrix $\mathcal{Q}$. Only the ${\tilde{h}_T}'$ matrices have to be updated each time, so we store an $RQ$ factorization of the $[G_B]_{s\times s} [\tilde{h}_{\Delta \tau}]_{s\times s}$ product for the duration of the update within a block for continuous-time/before a refresh for discrete-time. This $RQ$ factorization allows for the separation of scales, similar to the stabilization procedures used to compute Green's functions in AFQMC, and improves the precision for the determinants. It is important when there are many changes (and thus many matrices) in an updated block.

Finally, let us focus on the type-2 instability that occurs while recomputing the background matrix $G_B$ every time we move to the new spatial block in the same time interval $\Delta \tau$ in the continuous-time model update. This instability can also occur when we refresh $G_B$ within the same time-slice $t_\varepsilon$ in the discrete-time model. In both cases $G_B$ changes due to the changes in an $s\times s$ block in $h'_{\Delta \tau}$ or $h_{\tau_\varepsilon})$. Here we use the Woodbury identity to make the computational cost for the update to scale as $O(s N_s^2)$ while being careful about stability. Assuming $G_B = (1+\tilde{h}_B \tilde{h}_{\Delta \tau})^{-1} \tilde{h}_B \tilde{h}_{\Delta \tau}$ before the update, we compute the new $G'_B$ using the formula
\begin{equation}
\begin{aligned}
&\mathbbm{1}-{G_B}'\\
&=  (1+\tilde{h}_B \tilde{h}_{\Delta \tau}+\tilde{h}_B \tilde{h}_{\Delta\tau}(\tilde{h}_{\Delta\tau}^{-1} \tilde{h}'_{\Delta\tau} - \mathbbm{1}))^{-1}\\
& =  \mathbbm{1} - G_B \\
&\  + \big\{[G_B]_{N_s \times s} ([1-\mathcal{G}_{\Delta\tau}-G_B +2(\mathcal{G}_{\Delta \tau})G_B]_{s\times s})^{-1}\\
&\qquad\qquad\qquad\qquad\times[1-2\mathcal{G}_{\Delta\tau}]_{s\times s}[1-G_B]_{s\times N_s}\big\},
\label{woodbury}
\end{aligned}
\end{equation}
where we have defined the new Green's function within the time-slice
\begin{align}
\mathcal{G}_{\Delta\tau} = (\mathbbm{1}+ \tilde{h}_{\Delta\tau}^{-1} \tilde{h}'_{\Delta\tau})^{-1} \tilde{h}_{\Delta\tau}^{-1} \tilde{h}'_{\Delta \tau}.   
\end{align}
The symbols $[\;]_{s\times s}$, $[\;]_{N_s\times s}$ and $[\;]_{s \times N_s}$ stand for the appropriate blocks of matrices chosen.
This expression is in a slightly different form than the typical Woodbury formula \cite{doi:10.1137/1031049} and is written specifically in terms of $G_B$ and $\mathcal{G}_{\Delta\tau}$ because they are numerically stable to compute.

\begin{figure}
    \centering
    \includegraphics[width=8cm]{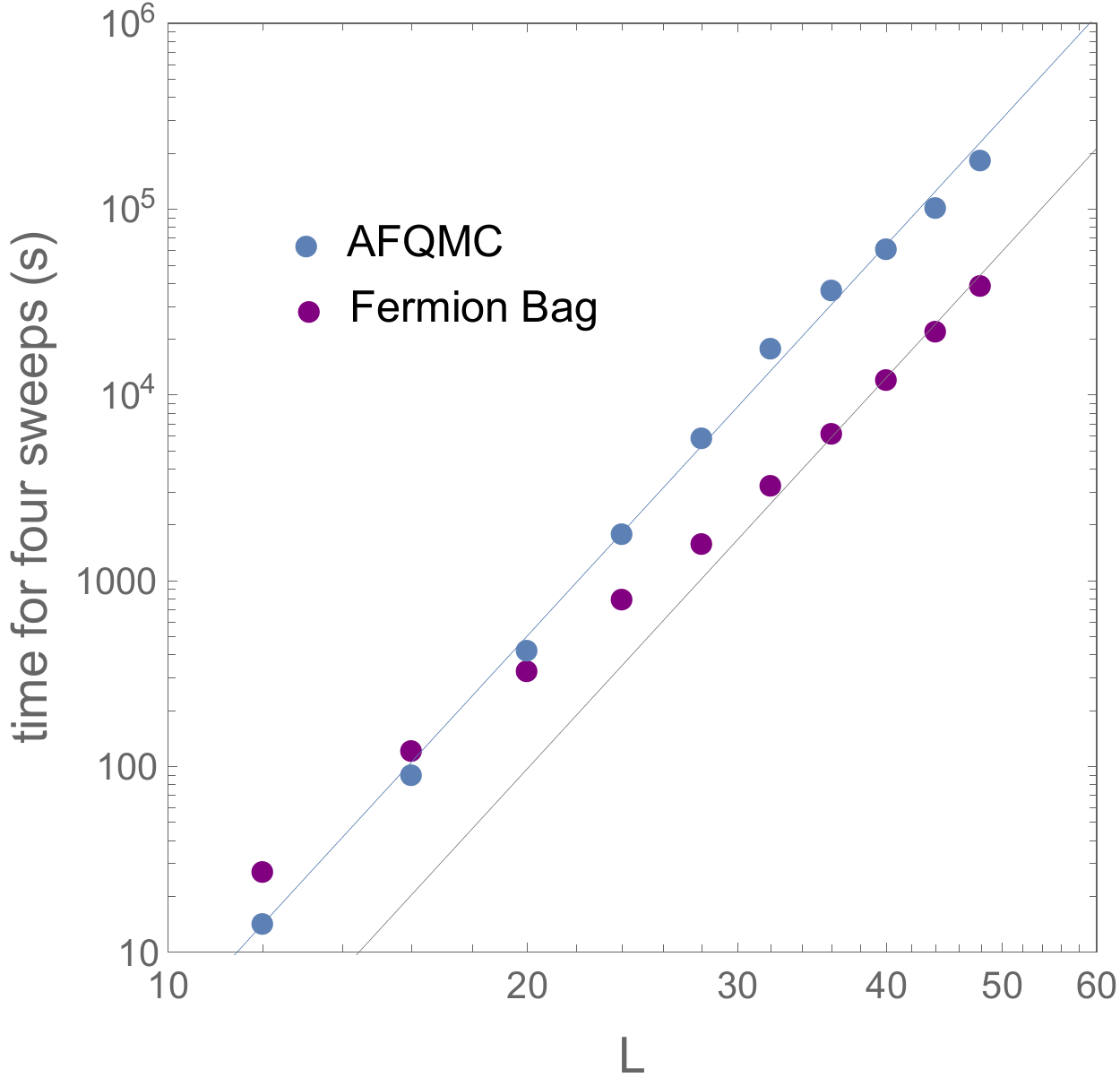}
    \caption{Scaling comparison for the Hamiltonian discrete-time fermion bag algorithm (Fermion Bag), and the auxiliary field algorithm (AFQMC) using the Algorithms for Lattice Fermions (ALF) software. The discrete time step is $\Delta \tau = 0.1$. Sweep times were measured for equilibrated configurations. The solid lines are $\tau_1 = 7.57467\times10^{-8} L^7 s$ for the fermion bag algorithm and $\tau_2 = 3.93116\times 10^{-7} L^7 s$ for
the auxiliary field algorithm.
}
    \label{timingcomparison}
\end{figure}



Using the idea of fermion bags can also help speed up the calculation at various stages in both continuous-time and discrete-time models.
For example, when we compute $\tilde{h}_{\Delta\tau}$, the calculations only involve multiplication of matrices within fermion bags associated with the region $\Delta \tau$ and they are all disconnected. This implies that if a fermion bag has $f$ sites associated with it, then we can construct the Greens function $G_f$ for that fermion bag alone as a matrix with $f$ rows and $f$ columns corresponding to the fermion bag sites. We can then combine the $G_f$ matrices for the different fermion bags into the $G_{\Delta \tau}$ matrix, which will naturally be block diagonal. So we can build $G_{\Delta \tau}$ rather quickly even on large spatial lattices. Thus, while $G_B$ is an $N_s\times N_s$ matrix, the concept of fermion bags along with the identity (\ref{buildg}) help reduce the number of operations needed.

As we build $G_B$ from the $G_{\Delta \tau}$ matrices, various steps can be stored either in computer memory or on the hard disk. This information allows us to make fast updates to $G_B$ when we move sequentially through the various $\Delta \tau$ regions by only having to combine one or two matrices per update of $G_{\Delta \tau}$. This allows us to keep the linear $\beta$ scaling. More details on how the storage scheme works can be found in the Appendix of \cite{Wang:2015rga}. The time to complete a single bond-update on the whole space-time lattice scales as $O(\beta N_s^3)$ for large system sizes. While this is identical to the traditional AFQMC method method, the idea of fermion bags significantly reduces the prefactor, as seen in Figure \ref{timingcomparison}, which we reproduce from the appendix of \cite{Huffman:2017swn}. This figure shows sweep times for the two methods using a $\Delta \tau = 0.1$. While these sweep times cannot be compared directly since they are not controlled for comparable precision in the observable, the plot illustrates how a significant number of the computations for the fermion bag algorithm, which would have normally had scaling of $O(\beta N_s^3)$, are instead replaced by computations with reduced scaling due to the use of the fermion bags. This is why the scaling looks to be smaller than $O(\beta N_s^3)$ for smaller lattices and then eventually goes to that expected scaling at large lattices, while for the AFQMC it is constant throughout.

\begin{table*}
\begin{center}
\begin{tabular}{|l||l|l|l|l|l|}
\hline
 $V/t$ & $L=20$ & $L=24$ & $L=32$ & $L=48$ & $L=64$ \\
\hline
$1.200$ & 0.00298(3) & 0.00184(3) &
0.00080(1) & $\qquad -$ & $\qquad -$ \\
\hline
$1.250$ & 0.00545(6) & 0.00380(5) & 0.00204(2) & 0.00074(2) & $\qquad -$ \\
\hline
$1.270$ & 0.00699(8) & 0.00517(7) & 0.00315(4) & 0.00151(3) &0.00085(1) \\
\hline
$1.280$ & 0.00787(9) & 0.00590(9) & 0.00377(4) & 0.00204(3) & 0.00130(2)  \\
\hline
$1.296$ & 0.00946(10) & 0.00740(9) 
& 0.00512(6) & 0.00339(5) & $\qquad -$ \\
\hline
$1.304$ & 0.01022(8) & 0.00844(9) 
& 0.00611(6) & 0.00423(5) & $\qquad -$ \\
\hline
$1.350$ & 0.01705(16)* & 0.01522(16)* 
& 0.01426(18)* & $\qquad -$ & $\qquad -$\\
\hline
$1.400$ & 0.02707(20)* & 0.02630(35)* & 0.02637(38)* & $\qquad -$ & $\qquad -$\\
\hline
\end{tabular}
\vspace{.5cm}
\caption{Results for the correlation function $\langle C\rangle$ defined in \cref{corrfn}, for the continuous-time model near the quantum critical point. A seven parameter fit of the data (after removing those marked with a "*") to the form \cref{scaleapprox} yields $\eta=.51(3)$, $\nu=.89(1)$, $V_c= 1.281(2)t$, $f_0=0.72(6)$, $f_1=0.29(2)$, $f_2=0.051(5)$, $f_3=0.0034(5)$, with a $\chi^2=0.90$.}
\label{datatable}
\end{center}
\end{table*}

\begin{table*}
\begin{center}
\small
\begin{tabular}{|l||l|l|l|l|l|l|l|}
\hline
 $V/t$ & $L=16$ & $L=20$ & $L=24$ & $L=32$ & $L=48$ & $L=64$ & $L=100$ \\
\hline
$1.00$ & 0.000217(6)* & 0.000100(2)* & 0.000048(1)* &
0.0000150(5)* & $\qquad -$ & $\qquad -$ & $\qquad -$\\
\hline
$1.36$ & 0.00237(3)* & 0.00159(3) & 0.00113(2) &
0.000598(2) & 0.000225(5) & 0.000117(6) & $\qquad -$\\
\hline
$1.38$ & 0.00276(4) & 0.00188(3) & 0.00134(2) & 0.000767(2) & 0.000318(6) & 0.000172(7) & $\qquad -$\\
\hline
$1.40$ & 0.00314(4) & 0.00215(3) & 0.00157(2) & 0.00096(2) & 0.000478(9) & 0.000271(8) & 0.000101(7) \\
\hline
$1.42$ & 0.00351(5) & 0.00256(4) & 0.00191(3) & 0.00127(2) & 0.00068(1) & 0.00042(2) & 0.00022(2) \\
\hline
$1.44$ & 0.00408(6) & 0.00304(4) & 0.00239(4) & 0.00167(3) & 0.00104(2) & 0.00079(2) &  $\qquad -$ \\
\hline
$1.50$ & 0.0061(1)* & 0.00495(7)* & 0.0042(1)* & 0.00397(8)* & 0.00404(8)* &  $\qquad -$ &  $\qquad -$ \\
\hline
\end{tabular}
\vspace{.5cm}
\caption{Results for the correlation function $\langle C\rangle$ defined in \cref{corrfn}, for the discrete-time model near the quantum critical point. A seven parameter fit of the data (after removing those marked with a "*") 
gives us $\eta=0.49(4)$, $\nu=0.94(3)$, $V_c= 1.420(2)t$, $f_0=0.22(2)$, $f_1=0.08(1)$, $f_2=0.013(3)$, $f_3=0.0010(3)$, with a $\chi^2=1.088$.}
\label{datatable2}
\end{center}
\end{table*}

\begin{figure*}
\begin{center}
\includegraphics[width=0.49\textwidth]{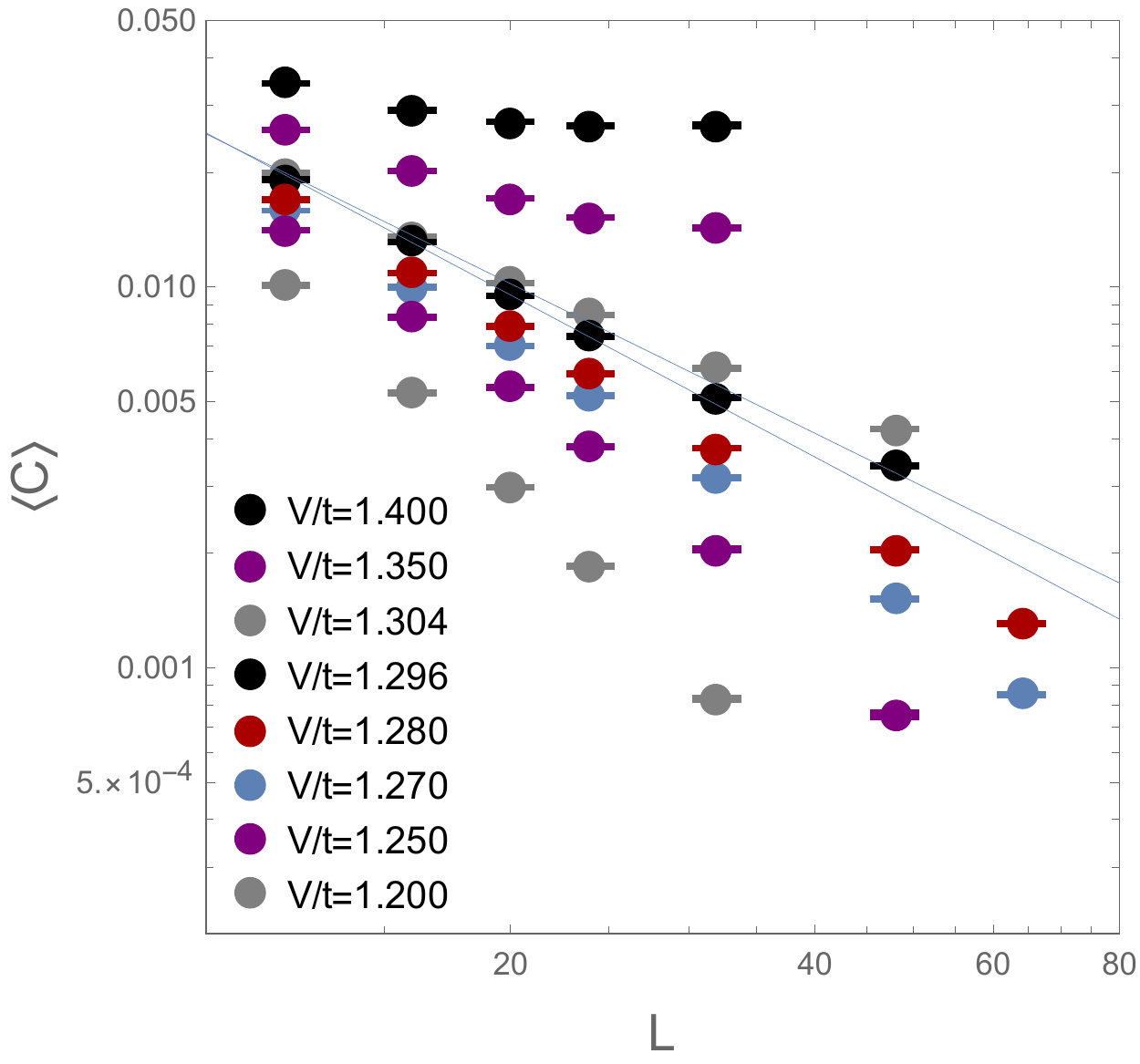}
\includegraphics[width=0.48\textwidth]{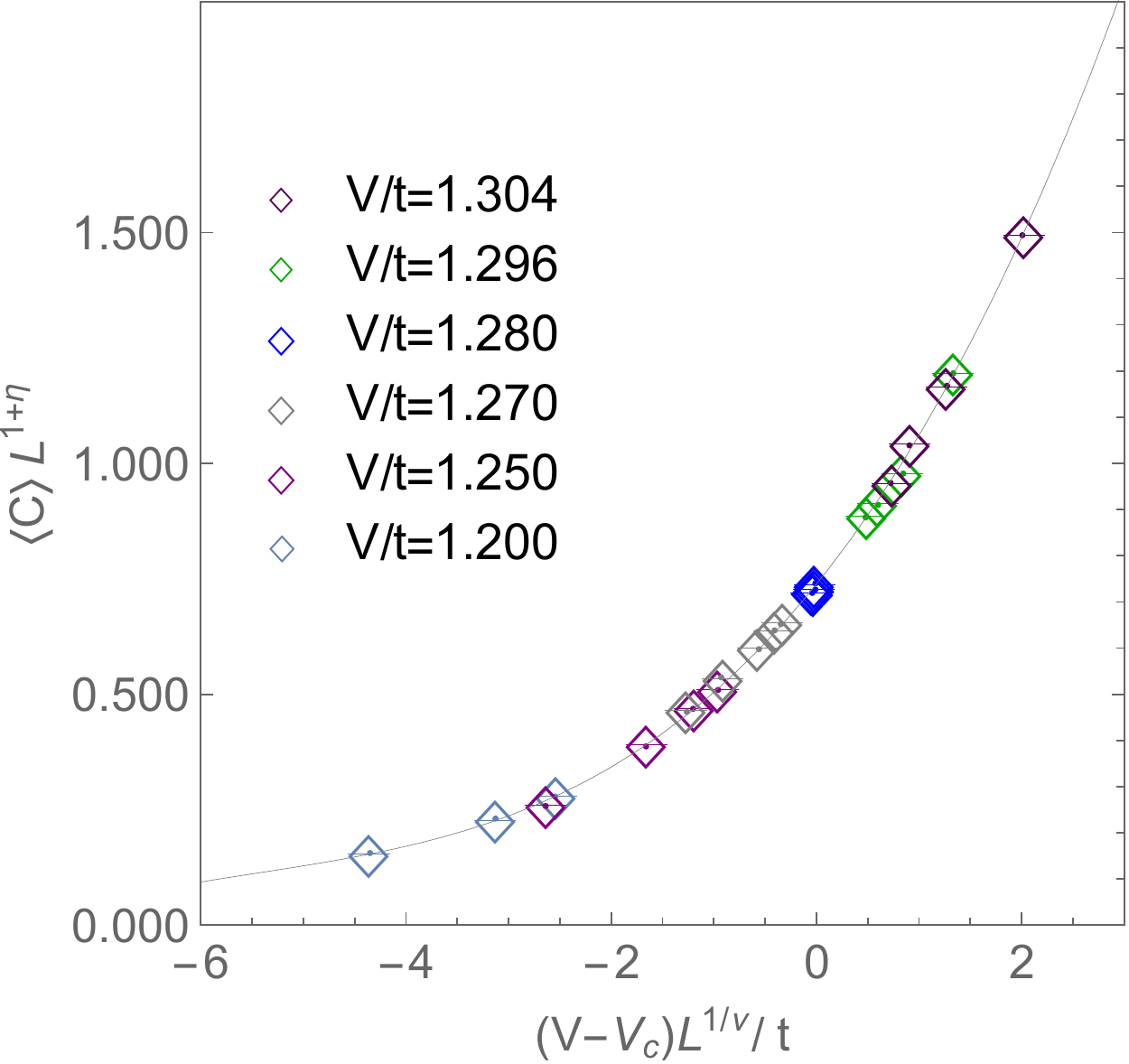}
  \end{center}
 \caption{The left plot shows $\langle C\rangle$ as a function of $L$ (with $\beta=L$) for different values of $V$ in the continuous-time model near the critical coupling. The solid lines show how the larger lattice data rules out $V=1.296t$ and $V=1.304$ as critical couplings, and both of these couplings are in fact in the broken phase. The right plot shows that most of the data displayed in the left plot collapses to a single critical scaling function. }
 \label{vcscals2}
\end{figure*}

\begin{figure*}
\begin{center}
\includegraphics[width=0.48\textwidth]{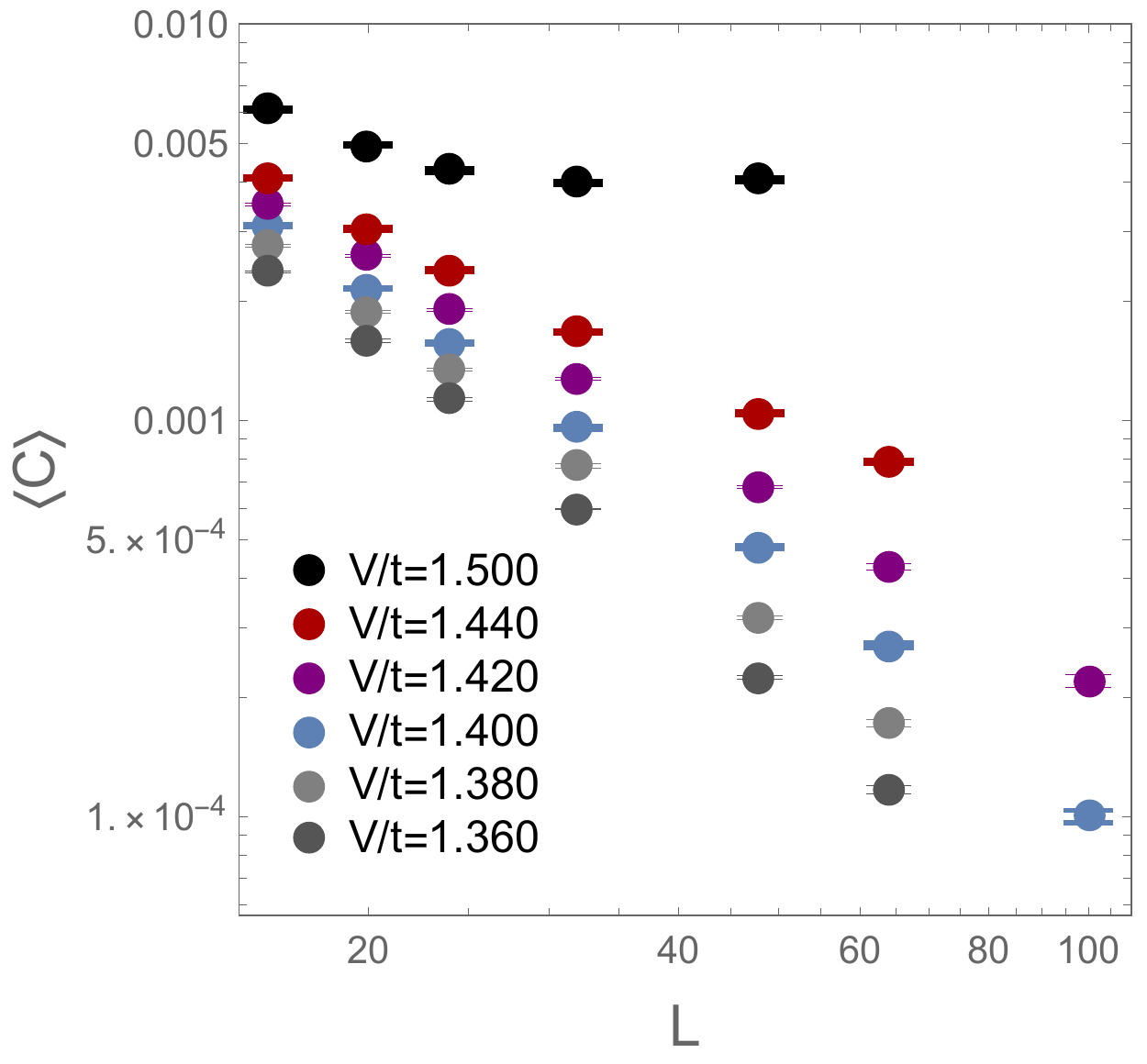}
\includegraphics[width=0.46\textwidth]{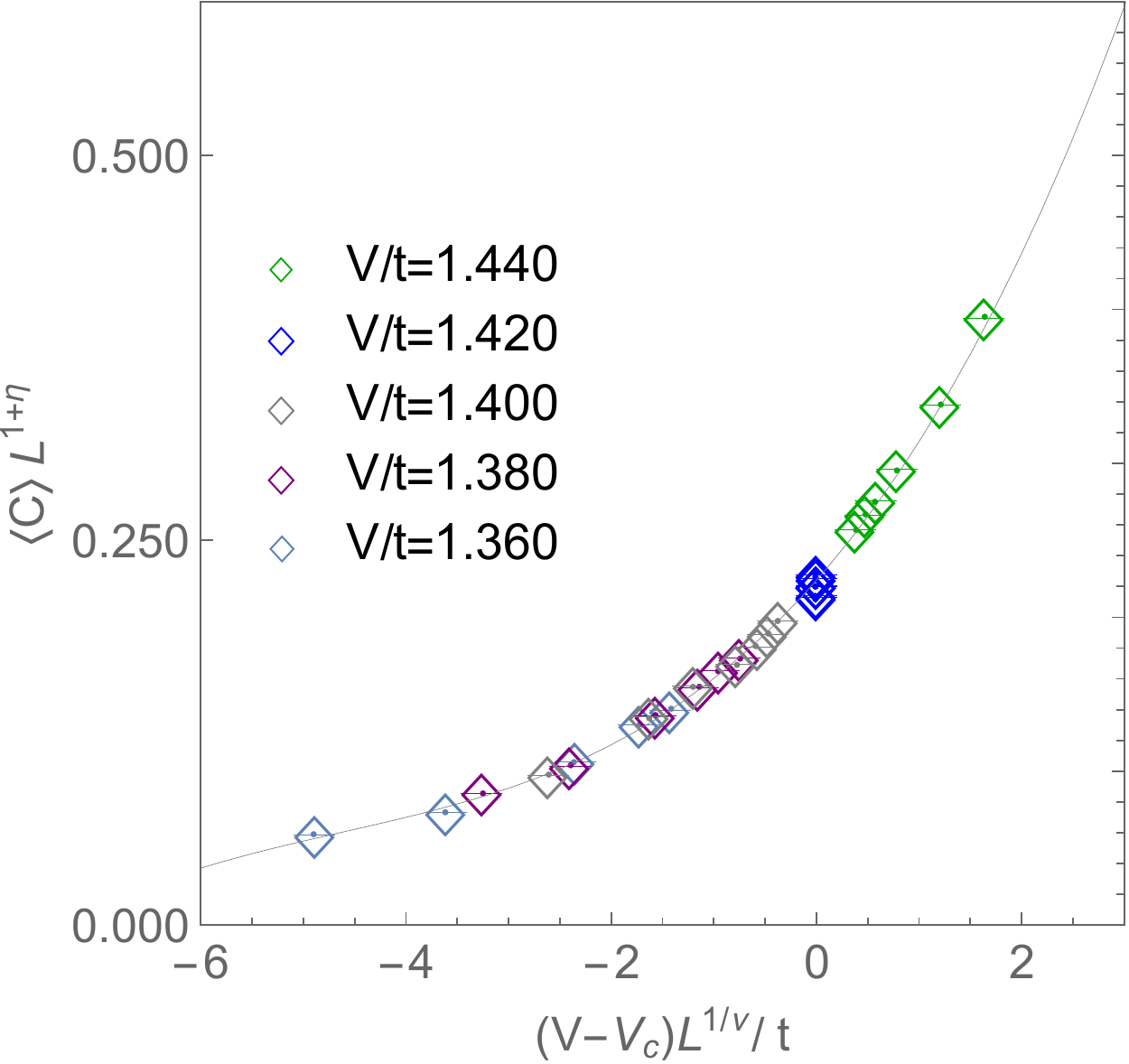}
\end{center}
\caption{The left plot shows $\langle C\rangle$ as a function of $L$ (with $\beta/\varepsilon = L$) for different values of $V$ in the discrete-time model near the critical coupling. The right plot shows that most of the data displayed in the left plot collapses to a single critical scaling function. }\label{vcscals3}
\end{figure*}

\section{Results}
\label{sec:results}

We now present results from our study of the model described by \cref{tvmodel} near the quantum critical point that separates the massless fermion phase at small values of $V/t$ from the massive fermion phase at large $V/t$ values. In the free limit, the theory describes massless Hamiltonian staggered fermions, which gives $N_f=1$ free massless Dirac fermions at long distances. Since the interaction term is invariant under translations by one lattice spacing and particle hole transformations, it forbids the staggered fermion mass term $(-1)^{x_1+x_2} (n_x-1/2)$ from being generated by radiative corrections. Since four-fermion interactions are also perturbatively irrelevant in $2+1$ dimensions, at small couplings the fermions remain massless. However, at large couplings the discrete symmetry breaks spontaneously and fermions become massive. So the long distance physics of the quantum critical point model is expected to be described by the $2+1$ dimensional $N_f=1$ Gross-Neveu \textit{chiral-Ising} universality class. 
 Since the non-zero value of $\varepsilon$ does not change the symmetries of the model, we expect that the temporal lattice spacing $\varepsilon$ does not change the universality class at least between $\varepsilon \rightarrow 0$ (continuous-time model) and $\varepsilon = 1$ (discrete-time model). Our main goal in this section is to confirm this.

Using the fermion bag algorithms described in the previous sections, we compute the two-point correlation function of the staggered mass order parameter $\langle C\rangle$ (see \cref{corrfn}) for both the models. Note that while the expectation value of the mass operator vanishes because of the Ising symmetry, the correlation function remains non-zero. For large lattice sizes $L$ we expect
\begin{equation}
\small
\left\langle C \right\rangle \sim \left\{ 
\begin{array}{ccc} L^{-4}&, & V < V_c \\
1&, & V > V_c\\
L^{-1-\eta}&, & V = V_c
\end{array} \right. ,
\label{piecescale}
\end{equation}
and near the critical point we expect the critical scaling relation
\begin{equation}
\langle C \rangle = \frac{1}{L^{1+\eta}} f\left(\left(V-V_c\right)L^{1/\nu}/t\right), \qquad V\approx V_c,
\label{scaleapprox}
\end{equation}
to hold. In order to obtain the critical exponents, we approximate $f(x)=f_0+f_1 x + f_2 x^2 + f_3 x^3$. Thus in the critical region we expect our data must be described by a seven parameter fit. 

Our results for the correlation function in the continuous-time model near the critical region are given in \cref{datatable} and plotted in \cref{vcscals2}. In these calculations we assume $\beta = L$ since the quantum critical point is expected to be relativistic. Excluding the data for $V/t=1.350$ and $1.400$, we are able to perform a seven parameter combined fit of the remaining data. If we only include the smaller lattices (those $L<64$ from Table \ref{datatable}), we obtain $\eta = 0.54(6)$ and $\nu = 0.88(2)$ with the critical coupling $V_c = 1.279(3)t$, as seen in \cite{Huffman:2017swn}. By including the larger lattices, we have been able to reduce the error for the critical exponents but find that they still remain in the same range as our smaller lattice calculations. The inclusion of larger lattices allows us to compute $\eta=0.51(3)$, $\nu=0.89(1)$, and $V_c=1.281(2) t$. The results for $L=64$, obtained from XSEDE resources \cite{xsede}, help us to reduce the error on the $\eta$ exponent and rule out a particularly high value for it. For example in \cref{vcscals2}, we note that at $V = 1.270 t$ the data seems to be described by a single power of the form $L^{-(1+\eta)}$ for lattice sizes from $L=20$ up to $L=48$. In fact a fit of the data to this form (leaving out the $L=64$ data), gives us $\eta=0.74(2)$ with a reasonable $\chi^2/DOF$. However, once we include the $L=64$ data we no longer get a good fit from which we are able to conclude that $V=1.270t$ is below the critical coupling.

Since the discrete time model with $\varepsilon=1$ is a microscopically different model, the critical value of $V/t$ where the phase transition occurs will be different. The results near this critical point are given in \cref{datatable2} and plotted in \cref{vcscals3}. Since the computations are faster, using XSEDE resources we can now compute correlation functions for lattices up to $L=100$ (with $\beta/\varepsilon = L$). The seven parameter fit (where we drop the points marked with an asterix in \cref{datatable2}) gives $V_c= 1.420(2)$, $\eta=.49(4)$ and $\nu=.94(3)$. As expected the critical coupling is different, but the critical exponents are consistent with those for the continuous time model. Interestingly we do not get a better precision.

\begin{center}
\begin{table}
\begin{center}
\begin{tabular*}{\linewidth}{@{\extracolsep{\fill}}|l|l|l|l|}
\hline
\hline
  Method & $\nu$ & $\eta$ & $N_s$ \\
\hline
\hline
 $4 - \epsilon$ \cite{Ihrig:2018hho} & 0.898(30) & 0.487(12) & -- \\

 FRG \cite{PhysRevB.94.245102}   & 0.93(1) & 0.55 & --
\\
 Large-N \cite{Gracey:1992,Gracey:1993kc}    & 0.938 & 0.509 & --
\\
 bootstrap \cite{Iliesiu2018}   & 1.32 & 0.544 & --
 \\
\hline
\hline
LCT-INT QMC \cite{Wang:2015rga}  & 0.80(3) & 0.30(2) & $2\times 18^2$
\\
LCT-INT QMC \cite{Hesselmann:2016tvh}  & 0.74(4) & 0.275(25) & $2\times 21^2$
\\
MQMC \cite{Li:2014tla}  & 0.77(3) & 0.45(2) & $2\times 24^2$ \\
SLAC QMC \cite{DSchmidt}   & 0.912(34) & -- & $32^2$\\
\hline
CT-FB QMC   & 0.89(1) & 0.51(3) & $64^2$\\
DT-FB QMC  & 0.94(3) & 0.49(4) & $100^2$
\\
\hline
\hline
\end{tabular*}
\\
\caption{Critical exponents $\nu$ and $\eta$ for the $N_f=1$ chiral-Ising Gross-Neveu universality class, obtained from various continuum-analytic and lattice-QMC methods. The last column gives the maximum spatial lattice sizes $N_s$ used in the QMC calculations. The last two rows give results from our current work.}
\label{tab:compare}
\end{center}
\end{table}
\end{center}

Recently, there have been a significant number of QMC studies of the quantum critical behavior in the lattice model we consider \cref{tvmodel} and other formulations \cite{WangNJP2014,LiNJP2015,Hesselmann:2016tvh,DSchmidt,Huffman:2017swn}. Our studies add to this growing literature, and in particular our results are among the largest lattices ever studied. Calculations of critical exponents of the associated $2+1$ dimensional $N_f=1$ Gross-Neveu \textit{chiral-Ising} universality class have also been performed over the years using continuum methods like large $N$, $\epsilon$-expansion, functional renormalization group (FRG) and bootstrap \cite{Gracey:1992,Gracey:1993kc,PhysRevB.94.245102,Iliesiu:2015qra,Zerf:2017zqi}. The most recent results are from a four loop calculation using the $4-\epsilon$ expansion \cite{Ihrig:2018hho}. Results for the critical exponents $\eta$ and $\nu$ obtained from all these methods are summarized \cref{tab:compare}. Note that our results are clearly consistent with the latest $4-\epsilon$ expansion results.

An important observation is that results from small lattice-QMC calculations are not compatible with continuum-analytic results. Our results also help us understand the reasons for the discrepancies. For example, if we assume that the critical point is at $V_c/t=1.296$ or $1.304$ as was found in some of the previous calculations on smaller lattices, and fit our continuous-time data to the form $L^{-(1+\eta)}$, after dropping larger values of $L$ we get $\eta=0.41(4)$ and $\eta=0.31(4)$ respectively (see left plot in Figure~\ref{vcscals2}). This agrees with the earlier results from smaller lattices, and the fits for the corresponding smaller lattices in our data are shown by the blue solid lines in the figure. However, it is clear that the fits fail dramatically if data from larger lattices $L=32$ and $L=48$ are included.

\section{Conclusions}
\label{sec:conc}
In this work we construct a new type of HLFT starting with a Hamiltonian that is written as a sum of exponential operators (see \cref{eq:hlocal}) involving nearest neighbor hopping terms. This form of the Hamiltonian is inspired by the fermion bag approach and can be used to study fermionic quantum critical behavior involving $N_f$ flavors of massless Dirac fermions interacting with four-fermion interactions. We concretely showed this by studying the critical behavior, which falls in the $N_f=1$ Gross-Neveu chiral-Ising universality. For the lattice Hamiltonian studied in this work we constructed a path integral with a temporal lattice spacing $\varepsilon$ and discussed how the continuous-time model ($\varepsilon \rightarrow 0$) and the discrete-time model ($\varepsilon = 1)$ emerge. We explained how the fermion bag algorithm is constructed in both these cases and how the idea of fermion bags can allow us to speed up calculations over the traditional AFQMC approach. By studying the finite size scaling of the order parameter correlation function, we showed that the quantum critical behavior in both these models belong to the same universality class. We further showed that it is feasible to study lattices with up to $N_s = 100\times 100$ lattice sites in the discrete-time model. The effort to study such a large lattice is roughly the same as the effort to study $N_s = 64\times 64$ sites in the continuous-time model. This can be partly attributed to fewer stabilization steps necessary in the discrete-time approach, but mainly attributed to a much smaller average number of bonds in equilibrated configurations for the large $\epsilon=1$. The discrete-time method also has the potential to practically work in a larger parameter range than the continuous-time method for models generically, due to fewer stabilization issues. On the other hand, fluctuations also seem larger in the discrete-time model for the lattice Hamiltonian that we studied, and one has to run longer to get comparable precision to that of the continuous-time model. The critical exponents we obtain in both cases are consistent with the recent four loop $4-\epsilon$ expansion results. 


\section*{Acknowledgments}
We would like to thank Fakher Assaad, Ribhu Kaul, Hanqing Liu, Michael Scherer, and Uwe-Jens Wiese for illuminating discussions. Additionally we would like to thank Fakher Assaad for sharing ALF data to compare discrete-time simulations and Michael Scherer for details on the $4-\epsilon$ expanision results. The work of SC is supported by the U.S. Department of Energy, Office of Science, Nuclear Physics program under Award Numbers DE-FG02-05ER41368. Research at Perimeter Institute
(EH) is supported by the Government of Canada through
the Department of Innovation, Science and Economic Development Canada and by the Province of Ontario through the Ministry of Research, Innovation and Science. This work used the resources provided by Extreme Science and Engineering Discovery Environment (XSEDE), which is supported by National Science Foundation grant number ACI-1548562.

\bibliographystyle{apsrev4-1}
\bibliography{refs,fqcp}
\onecolumngrid

\newpage
\appendix

\section*{Appendix: Gross-Neveu Universality Classes}

From a field theoretic perspective, Gross-Neveu universality classes can be best understood through Yukawa field theories, since then the relevant fixed point of interest become accessible in $4-\epsilon$-expansion \cite{Zerf:2017zqi}. There is a lot of literature on how these universality classes emerge and some of the recent discussions motivated from condensed matter lattice models can be found in Refs. \cite{PhysRevB.79.085116,PhysRevB.80.205319,PhysRevB.80.075432}. However, for easier reading of our paper we define the universality classes by discussing the symmetries explicitly here. For a more complete discussion we refer the reader to the previous literature.

Since we are interested in $2+1$ dimensions, we will constrain ourselves to two spatial dimensions with position ${\bf r}$ and define $N_f$ four-component Dirac field operators $\psi_i({\bf r}), i=1,2..,N_f$ and their Hermitian conjugates $\psi^\dagger_i({\bf r})\equiv (x,y)$. We will focus on models that do not break the $SU(N_f)$ flavor symmetries. In these models the five $4\times 4$ anticommuting Hermitian Dirac matrices, $\gamma_\mu, \mu = 0,1,2,3,5$ play an important role in our construction. We will assume they are normalized so that $\gamma_\mu^2 = \mathbbm{1}$. The free Dirac Hamiltonian with $N_f$ flavors of massless fermions is then given by
\begin{align}
H^f_0 \ =\ \int d^2 {\bf r}\ \sum_{i=1}^{N_f} \Big\{\psi^\dagger_i({\bf r}) 
\Big[(\gamma_0 \gamma_1)\partial_x +  (\gamma_0 \gamma_2)\partial_y \Big]\psi({\bf r})\Big\}
\end{align}
It is easy to verify that $H^f_0$ is invariant under "parity" transformations
\begin{align}
x\rightarrow -x, \quad y \rightarrow y,\quad &\psi_i({\bf r}) \rightarrow  \gamma_1\psi_i({\bf r}),\quad  
\psi^\dagger_i({\bf r}) \rightarrow \psi^\dagger_i({\bf r})\gamma_1,
\nonumber \\
x\rightarrow x,\quad y \rightarrow -y,\quad &\psi_i({\bf r}) \rightarrow  \gamma_2\psi_i({\bf r}),\quad 
\psi^\dagger_i({\bf r}) \rightarrow \psi^\dagger_i({\bf r})\gamma_2.
\end{align}
and the following $SU(2)$ transformations (which are referred to as "chiral transformations" although perhaps a misnomer)
\begin{align}
&\psi_i({\bf r}) \rightarrow  e^{i\theta_a S_a} \psi_i({\bf r}),\quad  
\psi^\dagger_i({\bf r}) \rightarrow \psi^\dagger_i({\bf r}) e^{-i\theta_a S_a}
\end{align}
where $S_a,a=1,2,3$ are three generators of $SU(2)$ transformations in the Dirac space given by
\begin{align}
S_1 = \frac{1}{2}\gamma_3, \quad S_2 = \frac{1}{2}\gamma_5,\quad S_3 = 
\frac{-i}{2}\gamma_3\gamma_5.
\end{align}
If we wanted to make the fermions massive, there are four possible Dirac mass terms which we write
\begin{align}
H^f_m \ =\ \int d^2 {\bf r}\ \sum_{i=1}^{N_f} \Big\{\psi^\dagger_i({\bf r}) (i\gamma_0\gamma_3\gamma_5)
\Big[m_0 \mathbbm{1}  + m_1 S_1 +  m_2 S_2 + m_3 S_3\Big]\psi({\bf r})\Big\},
\end{align}
out of which the terms multiplying $m_0, m_3$ break parity while the other two are parity invariant. On the other hand the $m_0$ term is invariant under chiral transformations while the other three terms transform as a 3-vector under the $SU(2)$ chiral transformations. To see this it is helpful to note that $(i\gamma_0\gamma_3\gamma_5)$ commutes with all $S_a$'s. Thus, we learn that it is possible to give the fermions in $2+1$ dimensions a chirally invariant mass term $m_0 \neq 0$ but such a term will break parity. On the other hand if we allow chiral symmetry breaking, we can give the fermions a parity invariant mass.

\begin{table}
\footnotesize
\begin{center}
\begin{tabular*}{.8\linewidth}{@{\extracolsep{\fill}}|p{3.5cm}|p{1.1cm}|p{1.1cm}|p{1.1cm}||p{2.6cm}|p{1cm}|p{1cm}|p{1cm}|}
\hline
\multicolumn{8}{|c|}{} \\
\multicolumn{8}{|c|}{\normalsize{\textbf{$N_f=1$ Four-Component Dirac Fermions}}} \\
\multicolumn{8}{|c|}{} \\
\hline
\hline
\multicolumn{4}{|c||}{} & \multicolumn{4}{c|}{}\\
\multicolumn{4}{|c||}{\small{$\mathbb{Z}_2$ Symmetry--\textit{Chiral Ising}}} & \multicolumn{4}{c|}{\small{$U(1)$ Symmetry--\textit{Chiral XY}}} \\
\multicolumn{4}{|c||}{} & \multicolumn{4}{c|}{}\\
\hline
  \textbf{Method} & $\nu$ & $\eta_\phi$ & $\eta_\psi$ & \textbf{Method} & $\nu$ & $\eta_\phi$ & $\eta_\psi$ \\
\hline
 $4 - \epsilon$ \textbf{(four loop with Borel \newline resummation)} \cite{Ihrig:2018hho} & 0.898(30) & 0.487(12) & 0.102(12) &  $4-\epsilon$ \textbf{(two loop)} \cite{Zerf:2017zqi,fei2016yukawa} & 0.883 & 0.574 & 0.25   \\
\hline
 \textbf{FRG} \cite{PhysRevB.94.245102}     & 0.93(1) & 0.5506 &  0.0654 & $4-\epsilon$ \textbf{(four loop)} \cite{Zerf:2017zqi} & 0.909 & 0.781 & 0.306 
\\
\hline
 \textbf{Large-N} \cite{Gracey:1992,Gracey:1993kc}    & 0.938 & 0.509 &  0.1056 & \textbf{FRG} \cite{PhysRevB.96.115132} & 1.21 & -- & -- 
\\
\hline
 \textbf{conformal bootstrap} \cite{Iliesiu2018}   & 1.32 & 0.544 &  0.084 & & & & 
 \\
\hline
\hline
  \textbf{CT-FB} \cite{Huffman:2017swn} \textbf{QMC (staggered fermions)} , \textit{here}  & 0.89(1) & 0.51(3) & -- & \multicolumn{4}{c|}{\small{$SU(2)$ Symmetry--\textit{Chiral Heisenberg}}}  \\
\hline
  \textbf{DT-FB} \textbf{QMC (staggered fermions)}, \textit{here}  & 0.94(3) & 0.49(4) & -- & \textbf{Method} & $\nu$ & $\eta_\phi$ & $\eta_\psi$
\\
\hline
  \textbf{HMC (SLAC fermions)} \cite{DSchmidt}   & 0.912(34) & -- & -- & $4-\epsilon$ \textbf{(one loop)} \cite{Zerf:2017zqi,Ros93}* & 0.900 & 0.667 & 0.5\\
  \hline
  \textbf{MQMC}\newline \textbf{ (honeycomb lattice)} \cite{Li:2014tla}  & 0.77(3) & 0.45(2) & -- & $4-\epsilon$ \textbf{(two loop)} \cite{Zerf:2017zqi} & 1.04 & 0.730 & 0.462
\\
\hline
  \textbf{MQMC}\newline \textbf{ (staggered fermions)} \cite{Li:2014tla}  & 0.79(4) & 0.43(2) & --& $4-\epsilon$ \textbf{(four loop)} \cite{Zerf:2017zqi} & 0.480 & 0.842 &  0.387
\\
\hline
  \textbf{LCT-INT QMC (honeycomb lattice)} \cite{Wang:2015rga}  & 0.80(3) & 0.302(7) & -- &\textbf{AFQMC} \textbf{(SLAC fermions)} \cite{Lang:2018csk} & 0.98(1) & 0.53(1) &  0.18(1)
\\
\hline
 \textbf{LCT-INT QMC (staggered fermions)} \cite{Wang:2015rga}  & 0.80(6) & 0.318(8) & -- & &  &  &  
\\
\hline
 \textbf{CT-INT QMC (honeycomb lattice)} \cite{Hesselmann:2016tvh}  & 0.74(4) & 0.275(25) & -- &  &  &  &  
\\
\hline
\end{tabular*}
\vskip0.15in
\begin{tabular*}{\linewidth}{@{\extracolsep{\fill}}|p{2.4cm}|p{1.1cm}|p{0.9cm}|p{0.9cm}||p{2cm}|p{.8cm}|p{.8cm}|p{.8cm}||p{1.8cm}|p{1cm}|p{1.1cm}|p{1cm}|}
\hline
\multicolumn{12}{|c|}{} \\
\multicolumn{12}{|c|}{\normalsize{\textbf{$N_f=2$ Four-Component Dirac Fermions}}} \\
\multicolumn{12}{|c|}{} \\
\hline
\hline
\multicolumn{4}{|c||}{} & 
\multicolumn{4}{c||}{} & 
\multicolumn{4}{c|}{} \\
\multicolumn{4}{|c||}{\small{$\mathbb{Z}_2$ Symmetry--\textit{Chiral Ising}}} & \multicolumn{4}{c||}{\small{$U(1)$ Symmetry--\textit{Chiral XY}}} & \multicolumn{4}{c|}{\small{$SU(2)$ Symmetry--\textit{Chiral Heisenberg}}} \\
\multicolumn{4}{|c||}{} & 
\multicolumn{4}{c||}{} & 
\multicolumn{4}{c|}{} \\
\hline
  \textbf{Method} & $\nu$ & $\eta_\phi$ & $\eta_\psi$ & \textbf{Method} & $\nu$ & $\eta_\phi$ & $\eta_\psi$ & \textbf{Method} & $\nu$ & $\eta_\phi$ & $\eta_\psi$ \\
\hline
 $4 - \epsilon$ / $2+\epsilon$ \newline \textbf{(four loop, Pad\'e approx., Borel resum.)} \cite{Ihrig:2018hho} & 1.01(3) & 0.72(2) & 0.043(1) & $4-\epsilon \quad P_{[2/2]}$ \textbf{(four loop)}  \cite{Zerf:2017zqi}  & 1.190 & 0.810 & 0.117 & $4-\epsilon \quad P_{[2/2]}$ \textbf{(four loop)}  \cite{Zerf:2017zqi}   & 1.5562 & 0.9985  &  0.1833  \\
\hline
   \textbf{Large-N} \cite{Gracey:1992,Gracey:1993kc,Kark94} &  1.050 & 0.743 &  0.044 & $4-\epsilon \quad P_{[3/1]}$ \textbf{(four loop)}  \cite{Zerf:2017zqi}  & 1.189 & 0.788 & 0.108  & $4-\epsilon \quad P_{[3/1]}$ \textbf{(four loop)}  \cite{Zerf:2017zqi}  & 1.2352 &  0.9563  & 0.1560
\\
\hline
   \textbf{FRG} \cite{PhysRevB.94.245102} & 1.006(2) &  0.7765 & 0.0276 & \textbf{Large-N} \cite{Li2017} &  1.25 & 0.67 & -- & \textbf{FRG} \cite{Knorr_2018} & 1.258 &  1.032 & 0.071
\\
\hline
   \textbf{conformal bootstrap} \cite{Iliesiu2018} & 1.14 & 0.742  &  0.044   & \textbf{FRG} \cite{PhysRevB.96.115132} & 1.160 & 0.88 & 0.062 & \textbf{FRG} \cite{PhysRevB.89.205403} & 1.31 & 1.02 & 0.08
 \\
\hline
\hline
  \textbf{Lagrangian FB QMC (staggered fermions)} \cite{PhysRevD.88.021701} & 0.83(1) & 0.62(1)  & 0.38(1)  & \textbf{AFQMC (honeycomb lattice)} \cite{Li2017} & 1.06(5) & 0.71(3) & -- & \textbf{HMC \newline (honeycomb lattice)}\cite{PhysRevB.98.235129} & 1.162 & 0.872(22) & --  \\
\hline
  \textbf{HMC (staggered fermions)} \cite{Kark94} & 1.00(4) & 0.754(8) & -- & \textbf{Lagrangian FB QMC\newline (staggered fermions)} \cite{PhysRevD.88.021701,PhysRevLett.108.140404} & 0.85(1) & 0.64(1) & 0.37(1) & \textbf{AFQMC (SM-QSH honeycomb lattice)}\cite{Liu2019}& 0.88(9) & 0.79(5) & -- 
\\
\hline
   \textbf{HMC 
   (SLAC fermions)} \cite{DSchmidt}  & 0.93(4) & -- & -- & \textbf{HMC (staggered fermions)} \cite{PhysRevB.79.241405} & 0.79(6) & 0.86(6) & -- & \textbf{AFQMC (staggered fermions /honeycomb lattice)} \cite{sorella} & 1.02(1) & 0.49(2) & 0.20(2) \\
   
   \hline
    \textbf{Designer model fermions+bosons QMC} \cite{Liu:2019xnb} & 1.0(1)  & 0.59(2)  & 0.05(2)  & \textbf{HMC (staggered fermions)} \cite{Armour:2009vj} & 0.87(3) & 0.64(3) & -- &  \textbf{AFQMC (staggered fermions /honeycomb lattice)} \cite{Parisen_Toldin_2015} & 0.84(4) & 0.70(15) & -- \\
       \hline
    \textbf{AFQMC (Dirac fermions+spins)}\cite{PhysRevB.97.081110} & 0.8(1)  & 0.65(3)  & --  &  &  &  &  &    &  &  &\\
  \hline
\end{tabular*}
\caption{Critical exponents according to continuum methods and QMC methods. *There is some disagreement over the counting of flavor numbers, as mentioned in \cite{sorella}, for the Chiral Heisenburg $4-\epsilon$ expansions. The number quoted for the (one-loop) calculation in \cite{Ros93} agrees with the number in \cite{Zerf:2017zqi}, but seems that it could be for one flavor of two-component fermions, instead of four-component fermions, according to the way the conventions appear in the work. Using those conventions, the numbers would instead be $\nu=0.882,\eta_\phi = 0.800,\eta_\psi=0.3$ for \cite{Ros93}, as given in \cite{sorella}. We have used more recent values for $4-\epsilon$ expansion-based results in other parts of table, based in \cite{Ihrig:2018hho,Zerf:2017zqi,fei2016yukawa}, but wanted to make clear this unresolved issue that could affect these Heisenberg class numbers.}
\label{tab:GNexp}
\end{center}
\end{table}

Gross-Neveu models preserve some subgroup of the $SU(2)$ chiral transformations and parity so that none of the mass terms are allowed. Hence these models can in principle be in two phases, either a massless fermion phase or a massive fermion phase that breaks either chiral symmetry or parity spontaneously. We will focus on those models that generate a parity invariant mass term. The phase transition between the two phases is then characterized by the Gross-Neveu universality class of the appropriate symmetry breaking pattern. Here we focus on three types of symmetry breaking pattern which are usually referred to in the literature as chiral-Ising, chiral-XY and chiral-Heisenberg. We construct Yukawa models for each of these three universality classes by coupling the free fermion theory with appropriate scalar fields.

\noindent {\em Chiral-Ising Symmetry:} This model preserves a $Z_2$ subgroup of the $SU(2)$ symmetry which then breaks to makes fermions massive. To construct the Yukawa model we introduce a single real scalar field $\phi({\bf r})$ and its canonical conjugate field $\pi({\bf r})$. The bosonic Hamiltonian $H_b(\phi,\pi)$ will be variant under the discrete symmetry transformation $\phi({\bf r}) \rightarrow -\phi({\bf r})$ and $\pi({\bf r}) \rightarrow -\pi({\bf r})$. The Hamiltonian of the Yukawa model is then given by
\begin{align}
H \ =\ H_b(\phi,\pi) + H^f_0 + g \int d^2 {\bf r}\ \sum_{i=1}^{N_f} \Big\{\psi^\dagger_i({\bf r}) (i\gamma_0\gamma_3\gamma_5) \ \big[\phi({\bf r}) S_3\big] \ \psi_i({\bf r})\Big\}
\end{align}
It is easy to verify that $H$ is invariant under discrete Ising subgroup of the $SU(2)$ chiral transformations
\begin{align}
\psi_i({\bf r}) \rightarrow  e^{i\pi S_1}\psi_i({\bf r}),\quad 
\psi^\dagger_i({\bf r}) \rightarrow \psi^\dagger_i({\bf r})e^{i\pi S_1},\quad
\phi({\bf r}) \rightarrow -\phi({\bf r}),\quad  \pi({\bf r}) \rightarrow -\pi({\bf r})
\end{align}
which means that a non-zero expectation value of the scalar field $\phi({\bf r})$ breaks this Ising symmetry and fermions can become massive.

\noindent {\em Chiral-XY Symmetry:} This model preserves a $U(1)$ subgroup of the $SU(2)$ symmetry which then breaks to makes fermions massive. To construct the Yukawa model we now introduce two real scalar fields $\phi_a({\bf r}),a=1,2$ and their canonically conjugate fields $\pi_a({\bf r})$. The bosonic Hamiltonian $H_b(\phi,\pi)$ will now be variant under the $U(1)$ symmetry transformations in which the complex fields $\phi({\bf r}) = \phi_1({\bf r}) + i\phi_2({\bf r})$ and $\pi({\bf r}) = \pi_1({\bf r}) + i\pi_2({\bf r})$ transform as $\phi({\bf r}) \rightarrow e^{-i\theta} \phi({\bf r})$ and $\pi({\bf r}) \rightarrow e^{-i\theta} \pi({\bf r})$. The Hamiltonian of the Yukawa model is then given by
\begin{align}
H \ =\ H_b(\phi,\pi) + H^f_0 + g \int d^2 {\bf r}\ \sum_{i=1}^{N_f} \Big\{\psi^\dagger_i({\bf r}) (i\gamma_0\gamma_3\gamma_5) \ \big[\phi_1({\bf r}) S_1 + \phi_2({\bf r}) S_2\big]\ \psi({\bf r})\Big\}.
\end{align}
It is easy to verify that $H$ is invariant under the $U(1)$ subgroup of the $SU(2)$ chiral transformations
\begin{align}
\psi_i({\bf r}) \rightarrow  e^{iS_3\theta} \psi_i({\bf r}),\quad 
\psi^\dagger_i({\bf r}) \rightarrow \psi^\dagger_i({\bf r})e^{-i\theta S_3},\quad
\phi({\bf r}) \rightarrow e^{-i\theta} \phi({\bf r}),\quad \pi({\bf r}) \rightarrow e^{-i\theta} \pi({\bf r}).
\end{align}
A non-zero expectation value of the complex scalar field $\phi({\bf r})$ then breaks the $U(1)$ chiral symmetry and thus fermions can become massive.

\noindent {\em Chiral-Heisenberg Symmetry:} Here we introduce three real scalar fields $\phi_a({\bf r}),a=1,2,3$ and their canonically conjugate fields $\pi_a({\bf r})$. The bosonic Hamiltonian $H_b(\phi,\pi)$ will now be variant under the $O(3)$ rotations of 3-vectors $\vec{\phi}({\bf r})$ and $\vec{\pi}({\bf r})$. The Hamiltonian of the Yukawa model is then given by
\begin{align}
H \ =\ H_b(\phi,\pi) + H^f_0 + g \int d^2 {\bf r}\ \sum_{i=1}^{N_f} \Big\{\psi^\dagger_i({\bf r}) (i\gamma_0\gamma_3\gamma_5)\ \big[\ \vec{\phi}({\bf r}) \cdot \vec{S}\ \big]\ \psi_i({\bf r})\Big\}.
\end{align}
It is easy to verify that $H$ is invariant under the full $SU(2)$ chiral transformations
\begin{align}
\psi_i({\bf r}) =  e^{i\theta \ \hat{n}\cdot \vec{S}} \psi_i({\bf r}),\quad 
\psi^\dagger_i({\bf r}) = \psi^\dagger_i({\bf r})e^{- i\theta \ \hat{n}\cdot \vec{S}},\quad
\vec{\phi}({\bf r}) \rightarrow R(\hat{n},\theta) \phi({\bf r}),\quad \vec{\pi}({\bf r}) \rightarrow R(\hat{n},\theta) \vec{\pi}({\bf r}),
\label{eq:chheis1}
\end{align}
where the $3\times 3$ matrix $R(\hat{n},\theta)$ rotates a 3-vector about the axis $\hat{n}$ by an angle $\theta$. A non-zero expectation value of the vector field $\vec{\phi}({\bf r})$ then breaks the $SU(2)$ chiral symmetry and makes fermions massive.

From the point of view of materials physics $N_f=2$ plays an important role due to the property that electrons have a spin-half and in many cases the $SU(2)$ spin symmetry can be treated as an internal symmetry. In such cases quantum phase transitions between massless and massive fermion phases can be described by other types of chiral Heisenberg models. For example consider the Hamiltonian given by
\begin{align}
H \ =\ H_b(\phi,\pi) + H^f_0 + g \int d^2 {\bf r}\ \sum_{i=1}^{N_f} \Big\{\psi^\dagger_i({\bf r}) (i\gamma_0\gamma_3\gamma_5)\ \big[\ \vec{\phi}({\bf r}) \cdot \vec{\sigma}\ \big]\ \psi_i({\bf r})\Big\}.
\label{eq:chheis2}
\end{align}
where now $\vec{\sigma}$ are Pauli matrices in the flavor space. In this case the massive fermion phase breaks flavor symmetry but preserves chiral symmetry. Such a Hamiltonian is expected to describe the Semi-Metal (SM) to Quantum-Spin-Hall (QSH) insulator transition as was recently studied in \cite{Liu2019}. Since for $N_f=2$, the $SU(2)$ chiral symmetry and $SU(2)$ flavor symmetries are equivalent \cref{eq:chheis1} and \cref{eq:chheis2} can be mapped into one another.

On the other hand most lattice formulations typically break the $SU(2)$ chiral symmetries to some subgroup, while preserving the $SU(2)$ flavor symmetries. Hence, one can look for other ways for generating mass terms with two flavors. For example the Hamiltonian
\begin{align}
H \ =\ H_b(\phi,\pi) + H^f_0 + g \int d^2 {\bf r}\ \sum_{i=1}^{N_f} \Big\{\psi^\dagger_i({\bf r}) (i\gamma_0\gamma_3\gamma_5)\ \big[\ \vec{\phi}({\bf r}) \cdot \vec{\sigma}\ S_3\ \big]\ \psi_i({\bf r})\Big\}.
\end{align}
preserves only discrete subgroup of the $SU(2)$ chiral symmetry but is $SU(2)$ flavor invariant. However, in the massive phase it breaks both the flavor and chiral symmetries. The phase transition in this model is expected to describe the Semi-Metal (SM) and an Anti-Ferromagnet (AFM) \cite{sorella,Parisen_Toldin_2015}. It has been suggested that the SM-QSH phase transition and SM-AFM transition could in fact belong to the same universality class \cite{Liu2019,PhysRevB.80.075432}.

Estimates for the critical exponents have been obtained from a variety of methods. While analytic methods can control the symmetries and symmetry breaking patterns, lattice QMC methods are less reliable due to fermion doubling problems and breaking of chira symmetries due to lattice artifacts. However, one can try to roughly count the fermion flavors and try to estimate the symmetry breaking patterns. Based on such estimates, in \cref{tab:GNexp} we tabulate the exponents obtained by various groups into the above three universality classes with $N_f=1$ and $N_f=2$.

\end{document}